\documentclass[reprint,amsmath,amssymb,aps,floatfix]{revtex4-2}
\usepackage{silence}
\usepackage{tikz}
\usepackage{quantikz}
\usepackage{graphicx}
\usepackage{subcaption}
\usepackage{threeparttable} 
\usepackage{dcolumn}
\usepackage{bm}
\usepackage{booktabs}   
\usepackage{bbold}
\usepackage{hyperref}
\usepackage{algorithm}
\usepackage{mathtools}
\usepackage{algpseudocode}
\usepackage[T1]{fontenc}
\usepackage{todonotes}
\begin{document}
\preprint{APS/123-QED}

\title{A Comparative Study of Quantum Optimization Techniques for Solving Combinatorial Optimization Benchmark Problems}

\author{Monit Sharma$^{1}$}
\author{Hoong Chuin Lau$^{1,2}$}
\email{Corresponding author email: hclau@smu.edu.sg}
\address{$^1$School of Computing and Information Systems,  Singapore Management University, Singapore}
\address{$^2$Institute of High Performance Computing, A*STAR, Singapore}

\begin{abstract}
Quantum optimization holds promise for addressing classically intractable combinatorial problems, yet a standardized framework for benchmarking its performance—particularly in terms of solution quality, computational speed, and scalability—is still lacking. In this work, we introduce a comprehensive benchmarking framework designed to systematically evaluate a range of quantum optimization techniques against well-established NP-hard combinatorial problems. Our framework focuses on key problem classes, including the Multi-Dimensional Knapsack Problem (MDKP), Maximum Independent Set (MIS), Quadratic Assignment Problem (QAP), and Market Share Problem (MSP)

Our study evaluates gate-based quantum approaches, including the Variational Quantum Eigensolver (VQE) and its CVaR-enhanced variant, alongside advanced quantum algorithms such as the Quantum Approximate Optimization Algorithm (QAOA) and its extensions. To address resource constraints, we incorporate qubit compression techniques like Pauli Correlation Encoding (PCE) and Quantum Random Access Optimization (QRAO). Experimental results, obtained from simulated quantum environments and classical solvers, provide key insights into feasibility, optimality gaps, and scalability. Our findings highlight both the promise and current limitations of quantum optimization, offering a structured pathway for future research and practical applications in quantum-enhanced decision-making.

\end{abstract}
\maketitle

\section{\label{sec:Intro}Introduction}

Combinatorial optimization plays a fundamental role in a wide range of scientific and industrial applications, including logistics \cite{yu2000robust}, finance \cite{zenios1993financial}, telecommunications \cite{resende2008handbook}, and drug discovery \cite{SINGH20221636}. Many critical problems in these fields fall within the class of NP-hard problems \cite{karp1972reducibility}, rendering them computationally intractable for large instances. Despite substantial advancements in classical optimization techniques, solving these problems remains challenging, as exact methods often become impractical due to their exponential scaling.

Quantum computing has emerged as a promising frontier in combinatorial optimization research, offering the potential to tackle these problems \cite{Abbas_2024}. The performance of quantum optimization algorithms is inherently tied to rigorous benchmarking. Unlike classical optimization, where well-established benchmarks exist \cite{classical_benchmark} for evaluating solvers, quantum optimization still lacks standardized methodologies and datasets that can reliably assess the effectiveness and scalability of quantum approaches. This gap hinders the ability to systematically compare algorithms, evaluate hardware capabilities, and understand the practical advantages quantum optimization might offer over classical techniques.

Benchmarking in quantum optimization serves multiple critical purposes. Firstly, it provides a structured way to measure progress in algorithmic and hardware development. The nascent field of quantum optimization is rapidly evolving, with new quantum algorithms being proposed, while the well known ones such as the Quantum Approximate Optimization Algorithm (QAOA) \cite{farhi2014quantumapproximateoptimizationalgorithm} and Variational Quantum Eigensolver (VQE) \cite{peruzzo2014variational}, are being refined \cite{Cerezo_2022,Zhang_2022,Blekos_2024}. Benchmarking enables researchers to evaluate these methods consistently across diverse problem instances and sizes, offering insights into their practical performance and limitations.

Secondly, benchmarking facilitates reproducibility and transparency in quantum computing research. With the inherent complexity of quantum hardware and its susceptibility to noise, it is crucial to establish robust benchmarks that can differentiate between genuine algorithmic improvements and artifacts caused by hardware-specific effects. Such benchmarks help create a level playing field for researchers and practitioners, fostering a collaborative environment for innovation.

Thirdly, benchmarks play a pivotal role in bridging the gap between theoretical promise and practical application. Many quantum algorithms exhibit theoretical advantages under specific assumptions, but their performance on real-world problems often deviates due to hardware constraints, such as limited qubit counts, decoherence, and gate fidelities. Benchmarking against practical datasets, such as those derived from real-world industrial problems, provides a reality check, highlighting the strengths and weaknesses of quantum optimization in practical settings.

The absence of a well-defined benchmarking framework also poses challenges for industry adoption. Decision-makers in industries such as logistics, finance, and supply chain management require clear evidence of the advantages of quantum optimization before committing resources to its implementation. A comprehensive benchmarking framework can provide this evidence, demonstrating the specific scenarios where quantum optimization outperforms classical methods or offers complementary benefits.

Moreover, benchmarking is essential for guiding the co-evolution of quantum hardware and algorithms. By identifying the problem instances and parameter regimes where quantum optimization excels, benchmarks inform hardware designers about the critical requirements for next-generation quantum devices. Similarly, they help algorithm developers tailor their methods to leverage the unique capabilities of quantum hardware.

To establish quantum optimization as a practical tool for solving real-world problems, the development of benchmarking frameworks must consider diverse factors. These include the selection of problem instances, the definition of performance metrics, and the handling of noise and errors in quantum computations. Additionally, benchmarking should address the trade-offs between solution quality, runtime, and resource utilization, providing a holistic view of quantum optimization's performance.

In this paper, we aim to bridge the existing gap by presenting a comprehensive benchmarking framework for quantum optimization. By tackling classically proven hard problems \cite{chekuri2004multidimensional, lawler1980generating, sahni1976p}, such as the Multi-Dimensional Knapsack Problem (MDKP) \cite{mdkp}, Maximum Independent Set (MIS) \cite{mis}, Market Share Problem (MSP) \cite{market_share} and Quadratic Assignment Problem (QAP) \cite{qap}, and using metrics tailored to quantum algorithms, we provide a robust foundation for evaluating quantum optimization methods. Our framework is designed to facilitate fair comparisons between classical and quantum solvers, highlight the unique strengths of quantum approaches, and identify areas where further research and development are needed.

These problems were selected (among many other candidates) because they not only represent classically proven hard problems that remain challenging for state-of-the-art classical solvers, but also give rise to real-world business problems (in logistics, supply chains and finance). Notably, QAP is among the ``hardest of the hard'' combinatorial optimization problems, as finding even an $\epsilon$-approximate solution has been proven to be NP-complete \cite{sahni1976p}. Moreover, other well-known NP-hard problems, such as the Traveling Salesman Problem (TSP), are special cases of QAP \cite{pardalos1994quadratic}.
Other challenging combinatorial optimization problems that were not considered in this study include the Low Autocorrelation Binary Sequences (LABS) problem \cite{packebusch2016low}, the Sports Timetabling problem \cite{van2023international}, among others.

\section{\label{sec:Contribution} Our Contributions}
In this paper, we make the following key contributions:

\begin{enumerate}
    \item \textbf{A Comprehensive Benchmarking Framework for Quantum Optimization:}  
    We present a rigorous benchmarking framework tailored for quantum optimization. This framework systematically evaluates a select set of combinatorial optimization problems that remain computationally challenging even at small scales. By focusing on problem instances that are within an intermediate size range where current and/or near-term quantum technologies can be effectively deployed, our approach enables a critical assessment of quantum techniques against current classical techniques.
    
    \item \textbf{Detailed Analysis of Hard Optimization Problems:}  
    We conduct a detailed examination of each benchmark problem, highlighting their inherent computational challenges and identifying key parameter settings that amplify their difficulty. This analysis not only explains why these problems remain intractable even at moderate scales but also serves as a valuable guide for selecting and tuning problem instances in future quantum optimization research.
    
    \item \textbf{Enhancement of the Market Share Problem:}  
    Among the four chosen problems, the Market Share Problem is the least well-studied in the classical optimization community. Hence, a side contribution of this work is to provide a formal problem formulation and a systematic approach for generating challenging test instances. Additionally, we revisit and significantly enhance classical CPLEX results for this problem \cite{market_share}, setting a new performance baseline for both classical and quantum optimization methods.
    
    \item \textbf{Advancements in Pauli Correlation Encoding (PCE):}  
    An emerging qubit-efficient technique is the Pauli Correlation Encoding (PCE) method \cite{sciorilli2025towards}. Another side contribution of this work is to refine PCE by introducing a QUBO-based loss function as an alternative to the conventional weighted Max-Cut formulation. Additionally, we implement a multiple re-optimization strategy that also incorporates an improved multi-step bit-swap operation as a post-processing technique, further enhancing solution quality.
    
    \item \textbf{Open-Source Accessibility:}
    Finally, in order to promote reproducibility and further research, we make all code, data, and benchmark instances publicly available in our GitHub repository \cite{Sharma2025QOB} and details on our quantum optimization algorithms can be found here \cite{Sharma2025QOA}.

\end{enumerate}

This paper is organized as follows: Section \ref{sec:Opt Problems} provides an overview of combinatorial optimization problems, discussing their common applications and practical significance. Section \ref{sec: qubo} introduces the QUBO formulation, explaining how integer linear programming (ILP) problems are transformed into QUBO models and exploring related formulations. Section \ref{sec:quantum_opto_algo} examines quantum optimization and its associated algorithms, highlighting the advantages of quantum computing and summarizing the available techniques.

Section \ref{sec:Problems} describes the benchmark problems used in this study, detailing their selection criteria and formal definitions. Section \ref{sec:level7} outlines the experimental setup, including hardware specifications and evaluation metrics. Section \ref{sec:level8} presents the results and analysis, while Section \ref{sec:level9} concludes the paper.
\section{\label{sec:Opt Problems} Combinatorial Optimization Problems}

Combinatorial optimization is a key topic in Operations Research, Computer Science and Applied Mathematics, driving decision-making across fields such as engineering, business, and computational sciences. It involves determining the optimal (minimum or maximum) value of an objective function while adhering to constraints that represent resource limitations, system dynamics, or problem-specific conditions.

The choice of an appropriate solution method depends on factors such as problem size, constraints, computational feasibility, and whether an exact or approximate solution is required. While classical techniques remain dominant for well-structured problems, emerging approaches—such as quantum and machine learning-based optimization—are becoming increasingly relevant for tackling high-dimensional, combinatorial, and non-convex problems.

Building on the detailed classification of classical optimization techniques, it is essential to explore emerging paradigms that extend beyond conventional computational approaches. Quantum optimization introduces a fundamentally different framework for solving combinatorial and continuous optimization problems by leveraging quantum mechanical principles. Unlike classical methods, which rely on iterative search heuristics or mathematical relaxations, quantum algorithms exploit superposition, entanglement, and interference to explore solution spaces more efficiently, particularly for problems characterized by complex landscapes and exponential search spaces.

Recent advancements in quantum hardware and algorithm development—along with the promise of future breakthroughs \cite{ibm_quantum_roadmap, quantinuum_roadmap_2024}—have heightened interest in assessing the viability of quantum optimization for solving real-world problems. While classical solvers remain the preferred choice for structured and well-behaved problems, quantum optimization techniques are actively being explored for their potential speedups and advantages in specific problem domains.

\section{\label{sec: qubo}QUBO Formulation}

Integer Linear Programming (ILP) involves optimizing a linear objective function subject to linear equality and inequality constraints, with variables restricted to integer values. To leverage quantum optimization techniques, particularly those designed for Quadratic Unconstrained Binary Optimization (QUBO) problems, it is essential to transform ILP formulations into QUBO representations. This transformation enables the application of quantum algorithms, such as quantum annealing, to solve problems originally expressed as ILPs.

Since most quantum optimization approaches, including those based on gate-based quantum computing and quantum annealing, are naturally suited for QUBO formulations, understanding the QUBO framework is fundamental for quantum algorithm design. Many combinatorial optimization problems, can be expressed in QUBO form, making it a widely used representation in quantum optimization research.

\subsection{Mathematical Formulation of ILP}

\subsubsection{General Form of ILP}

An ILP can be expressed as:

\begin{align}
    \text{Minimize} \quad & c^T x \\
    \text{Subject to} \quad & A_{\text{eq}} x = b_{\text{eq}} \\
    & A_{\text{ineq}} x \leq b_{\text{ineq}} \\
    & x \in \mathbb{Z}^n
\end{align}

where \( x \) is an \( n \)-dimensional vector of integer variables, \( c \) is a coefficient vector for the objective function, \( A_{\text{eq}} \) and \( A_{\text{ineq}} \) are matrices defining the equality and inequality constraints, and \( b_{\text{eq}} \) and \( b_{\text{ineq}} \) are corresponding constant vectors.

\subsection{Transformation to QUBO}

\subsubsection{Binary Encoding of Integer Variables}

Each integer variable \( x_i \) is represented using binary variables. If \( x_i \) has an upper bound \( U_i \), it can be expressed as:

\begin{equation}
    x_i = \sum_{k=0}^{K_i - 1} 2^k \, \psi_{ki}
\end{equation}

where \( \psi_{ki} \in \{0, 1\} \) are binary variables, and \( K_i = \lceil \log_2(U_i + 1) \rceil \) is the number of binary variables required.

\subsubsection{Reformulating the Objective Function}

Substituting the binary representations of the integer variables into the original objective function:

\begin{equation}
    c^T x = \sum_{i=1}^n c_i x_i = \sum_{i=1}^n c_i \sum_{k=0}^{K_i - 1} 2^k \, \psi_{ki}
\end{equation}

This results in a linear function of binary variables.

\subsubsection{Transforming Constraints}

For equality constraints $( A_{\text{eq}} x = b_{\text{eq}} $, substitute the binary representations of $ x $:

\begin{equation}
    \sum_{j=1}^n A_{\text{eq},ij} x_j = b_{\text{eq},i} \quad \Rightarrow \quad \sum_{j=1}^n A_{\text{eq},ij} \sum_{k=0}^{K_j - 1} 2^k \, \psi_{kj} = b_{\text{eq},i}
\end{equation}

To incorporate these constraints into the QUBO framework, add penalty terms to the objective function:

\begin{equation}
    P_{\text{eq}} = \lambda_{\text{eq}} \sum_{i=1}^{m_{\text{eq}}} \left( \sum_{j=1}^n A_{\text{eq},ij} \sum_{k=0}^{K_j - 1} 2^k \, \psi_{kj} - b_{\text{eq},i} \right)^2
\end{equation}

where \( \lambda_{\text{eq}} \) is a penalty coefficient, and \( m_{\text{eq}} \) is the number of equality constraints.

For inequality constraints \( A_{\text{ineq}} x \leq b_{\text{ineq}} \), introduce non-negative slack variables \( s_i \) to convert them into equalities:

\begin{align}
    \sum_{j=1}^n A_{\text{ineq},ij} x_j + s_i &= b_{\text{ineq},i} \\
    \Rightarrow \quad \sum_{j=1}^n A_{\text{ineq},ij} \sum_{k=0}^{K_j - 1} 2^k \, \psi_{kj} + s_i &= b_{\text{ineq},i}
\end{align}

Each slack variable \( s_i \) is also expressed in binary form:

\begin{equation}
    s_i = \sum_{l=0}^{L_i - 1} 2^l \, \phi_{li}
\end{equation}

where \( \phi_{li} \in \{0, 1\} \) are binary variables, and \( L_i = \lceil \log_2(S_i + 1) \rceil \), with \( S_i \) being an upper bound on \( s_i \).

Incorporate these transformed constraints into the objective function with penalty terms:

\begin{align}
    P_{\text{ineq}} = \lambda_{\text{ineq}} \sum_{i=1}^{m_{\text{ineq}}} 
    \bigg( &\sum_{j=1}^n A_{\text{ineq},ij} \sum_{k=0}^{K_j - 1} 2^k \, \psi_{kj} \notag \\
    &+ \sum_{l=0}^{L_i - 1} 2^l \, \phi_{li} - b_{\text{ineq},i} \bigg)^2
\end{align}

where \( \lambda_{\text{ineq}} \) is a penalty coefficient, and \( m_{\text{ineq}} \) is the number of inequality constraints.

\subsubsection{Constructing the QUBO Objective Function}

The final QUBO formulation consists of the transformed objective function combined with the penalty terms:

\begin{equation}
    Q(\psi, \phi) = \sum_{i=1}^n c_i \sum_{k=0}^{K_i - 1} 2^k \, \psi_{ki} + P_{\text{eq}} + P_{\text{ineq}}
\end{equation}

This results in a quadratic function of the binary variables \( \psi_{ki} \) and \( \phi_{li} \), suitable for QUBO solvers.

\subsection{Choosing Penalty Coefficients}

Selecting appropriate values for the penalty coefficients \( \lambda_{\text{eq}} \) and \( \lambda_{\text{ineq}} \) is crucial. If these coefficients are too small, the optimization process may yield solutions that violate the constraints. Conversely, excessively large penalties can dominate the objective function, potentially leading to numerical instability and poor convergence.

A practical approach is to set the penalty coefficients sufficiently high to ensure constraint satisfaction. One heuristic for selecting these coefficients is:

\begin{equation}
    \lambda_{\text{eq}}, \lambda_{\text{ineq}} > C_{\max}
\end{equation}

where \( C_{\max} \) is the maximum absolute value of the objective function coefficients \( c_i \). This ensures that any constraint violation is penalized more heavily than the contribution of any single term in the objective function. In practice, empirical tuning or adaptive penalty methods may be employed to achieve an optimal balance.

\subsection{Alternative QUBO Formulations}

While traditional QUBO formulations often incorporate slack variables to handle inequality constraints, recent research has proposed alternative methods that avoid the need for these additional variables. Few such approaches are the \textit{unbalanced penalization} \cite{montanez2024unbalanced} technique which directly integrates inequality constraints into the objective function using asymmetric penalty terms and the \textit{ subgradient method} \cite{takabayashi2024subgradientmethodusingquantum} that uses a Hubbard-Stratonovich transformation to relax equality constraints. In this paper, we stick with the slack-based formulation.

\section{\label{sec:quantum_opto_algo}Quantum Optimization Algorithms}

Classical methods, such as integer programming and enumerative techniques, often struggle with increasing complexity as problem sizes grow. In contrast, quantum algorithms use different computational principles that may help address scalability challenges, especially in large, high-dimensional, and combinatorial problems. Several quantum paradigms have been developed to tackle optimization tasks, leveraging distinct algorithmic strategies and hardware implementations. These methods range from variational quantum computing approaches to quantum annealing, quantum phase estimation, and adaptations of classical algorithms to quantum settings.

\subsection{Variational and Qubit-Efficient Quantum Algorithms}

Gate-based quantum computing employs a circuit where quantum states encode optimization problems and are manipulated using quantum gates. A key class of algorithms in this paradigm is \textbf{Variational Quantum Algorithms (VQAs)}, which reformulate optimization tasks as energy minimization problems. These algorithms leverage parameterized quantum circuits optimized through classical techniques.

A fundamental example of this approach is the \textbf{Quantum Approximate Optimization Algorithm (QAOA)} \cite{farhi2014quantumapproximateoptimizationalgorithm}, which alternates between quantum evolution and classical optimization to solve combinatorial problems. Several enhancements to QAOA have been developed, including, \textbf{Quantum Alternating Operator Ansatz (QAOAz)} \cite{hadfield2019quantum}, which generalizes alternating operators to enable richer solution space exploration, \textbf{Warm-Start QAOA (WS-QAOA)} \cite{Egger_2021}, which utilizes classical preprocessing to initialize QAOA parameters, accelerating convergence and \textbf{Multi-Angle QAOA (MA-QAOA)} \cite{herrman2022multi}, which introduces independent variational parameters for each layer, improving performance.

Another significant VQA is the \textbf{Variational Quantum Eigensolver (VQE)} \cite{peruzzo2014variational}, originally developed for quantum chemistry but extended to broader optimization problems. Enhancements like \textbf{Subspace-Search VQE (SSVQE)} \cite{Nakanishi_2019} and \textbf{Variational Quantum Deflation (VQD)} \cite{Higgott_2019} further improve the performance of VQE, particularly in multi-solution scenarios. Additionally, \textbf{CVaR VQE} \cite{Barkoutsos_2020} employs Conditional Value-at-Risk (CVaR) aggregation to refine convergence and solution quality by focusing on the lowest-energy measurement outcomes, i.e. instead of using the expectation value, it considers only the lowest fraction (confidence level $\alpha$) of the measured energies.

Scalability remains a major challenge in quantum optimization due to high qubit requirements. To address this, qubit-efficient representations and Reduction techniques have been explored. \textbf{Quantum Random Access Optimization (QRAO)} \cite{fuller2024approximate} compresses multiple binary variables into a single qubit, while \textbf{Pauli Correlation Encoding (PCE)} \cite{sciorilli2025towards} embeds problem constraints into Pauli correlations, reducing qubit overhead while maintaining solution accuracy. Prior research \cite{Sharma_2024} has explored solving constrained optimization problems using QRAO.

Hybrid quantum-classical methodologies are particularly well-suited for the \textit{Noisy Intermediate-Scale Quantum (NISQ)} era \cite{preskill2018quantum}, where fully fault-tolerant quantum computing is not yet available. By minimizing quantum resource demands, these methods enable practical quantum-assisted optimization while mitigating limitations such as short coherence times and gate errors.

\subsection{Adiabatic and Quantum-Enhanced Classical Techniques}

Beyond variational quantum algorithms (which is the focus of this paper), we also list other quantum paradigms that offer alternative approaches to solving optimization problems efficiently. One such approach is \textbf{Quantum Annealing and Adiabatic Quantum Computation}, which does not rely on gate operations. Instead, it employs a continuous evolution of system parameters, gradually transforming an initial quantum state into one that encodes the optimization problem’s solution. This method is particularly effective for problems formulated as QUBO or Ising models \cite{kadowaki1998quantum}.

The \textbf{Quantum Adiabatic Algorithm (QAA)} \cite{farhi2000quantum} extends this principle by leveraging the adiabatic theorem to ensure the system remains in the ground state during evolution. Variants such as \textbf{diabatic quantum annealing} \cite{crosson2021prospects} and \textbf{counter-diabatic annealing} \cite{del2013shortcuts} enhance efficiency by either relaxing the strict adiabatic condition or introducing additional Hamiltonian terms to improve performance.

Another critical quantum tool in optimization is \textbf{Quantum Phase Estimation (QPE)} \cite{kitaev1995quantum}, which extracts eigenvalues of unitary operators and serves as a fundamental building block for several optimization tasks. QPE plays a crucial role in \textbf{Quantum Amplitude Estimation (QAE)} \cite{brassard2002quantum}, which enables efficient sampling in quantum-enhanced optimization frameworks such as \textbf{Quantum Simulation-Based Optimization (QSBO)} \cite{gacon2020quantum, sharma2024quantum, sharma2024quantummontecarlomethods}.

Quantum optimization continues to evolve rapidly, with hybrid quantum-classical algorithms currently leading practical implementations. While scalability remains a pressing challenge, advancements in qubit-efficient representations and algorithmic improvements push the boundaries of computational feasibility. As quantum hardware advances, previously intractable optimization problems may become solvable, bridging the gap between theoretical breakthroughs and real-world applications.

\section{\label{sec:Problems}Benchmark Problems}

Identifying real-world optimization problems that remain computationally challenging even for relatively small instance sizes is essential for benchmarking quantum algorithms. Many commonly studied optimization problems are either synthetically generated or have well-established heuristics that scale efficiently in practice. However, finding \textbf{real-world instances} that are both computationally hard and relevant for practical applications remains a significant challenge.

In this work, we focus on problems that are not only NP-hard but also pose challenges in finding \textbf{either an optimal or even a feasible solution} within reasonable time constraints. While some problems are difficult due to the complexity of reaching optimality, others are hard because many potential solutions are infeasible.
Moreover, these problems encompass \textbf{varied objectives}, including feasibility determination, minimization, and maximization. 
Each problem can be formulated as a \textbf{Quadratic Unconstrained Binary Optimization (QUBO)} problem, ensuring compatibility with quantum optimization techniques.

Each benchmark problem instance is converted into a QUBO model, thereby standardizing the objective to minimization, which is essential for compatibility with our quantum optimization techniques. We employ a slack-based formulation for generating the underlying QUBO, as discussed in the section on QUBO Formulation above. The QUBO model is then solved using the various quantum methods, and the resulting solution is translated back to yield the solution for the original problem. 

In the following subsections, we provide mathematical formulation of each benchmark problem, outlining their computational challenges and the specific instances used for benchmarking.

\subsection{Summary of Problem Descriptions}

The following gives a summary of the problem statement of the problems considered:

\begin{enumerate}
    \item \textbf{Multi-Dimensional Knapsack Problem (MDKP)}: Select items with multiple attributes that maximizes total profit while satisfying multiple capacity constraints.
    \item \textbf{Maximum Independent Set (MIS)}: Find a largest subset of non-adjacent vertices in a given graph.
    \item \textbf{Quadratic Assignment Problem (QAP)}: Assign facilities to locations to minimize a cost function which is a sum product of distances and flows.
    \item \textbf{Market Share Problem (MSP)}: Assign customers or retailers to divisions to achieve a target market share distribution while minimizing deviations when an exact split is not feasible.
\end{enumerate}

\subsection{Benchmark Instances}

To effectively assess the performance of quantum optimization methods, we carefully select benchmark instances that are known to be classically hard to solve computationally. These instances exhibit characteristics such as:

\begin{itemize}
    \item \textbf{Strong dependencies between decision variables}, making heuristic approaches ineffective.
    \item \textbf{Scalability challenges}, where classical solvers struggle even for moderately sized instances.
    \item Each of the selected problem instances ensures at least one \textbf{feasible solution}. 
\end{itemize}

In the subsequent sections, we provide a detailed examination of each problem type, highlighting their computational difficulty, the specific benchmark instances used, and their relevance to real-world applications.

\subsection{Multi Dimensional Knapsack Problem }

The \textit{Multi-Dimensional Knapsack Problem} (MDKP) \cite{mdkp} is a generalization of the classical knapsack problem, widely studied in Operational Research. The problem involves selecting a subset of items to maximize the total profit while satisfying multiple resource constraints. It is widely used in multi-resource allocation problems such as project selection, budget planning, and inventory management.

\subsubsection{Mathematical Formulation}
The MDKP can be represented as the following integer program. Given a set of $n$ items and $m$ resource dimensions, the goal is to maximize the profit while ensuring that the total weight of selected items does not exceed the capacity of any resource dimension:
\begin{align*}
    \text{Maximize} \quad & \sum_{i=1}^n p_i x_i \\
    \text{subject to} \quad & \sum_{i=1}^n w_{ij} x_i \leq c_j, \quad \forall j \in \{1, \dots, m\}, \\
    & x_i \in \{0, 1\}, \quad \forall i \in \{1, \dots, n\}.
\end{align*}
where:
\begin{itemize}
    \item \( x_i \) is a binary decision variable, where \( x_i = 1 \) if item \( i \) is selected, and \( x_i = 0 \) otherwise.
    \item \( p_i \) represents the profit associated with item \( i \).
    \item \( w_{ij} \) denotes the weight of item \( i \) in dimension \( j \).
    \item \( c_j \) is the capacity limit for resource dimension \( j \).
\end{itemize}

\subsubsection{Factors Affecting Problem Complexity and Hardness}
The difficulty of solving the MDKP is influenced by several factors, which can be adjusted to generate harder instances:

\begin{enumerate}
    \item \textbf{Number of Constraints (\(m\))}: Increasing the number of constraints restricts the feasible solution space, making it more challenging to identify optimal solutions. Each additional constraint introduces an extra limitation on item selection, leading to higher computational complexity.
    
    \item \textbf{Correlation Between Weights and Profits}: Instances where item weights (\( w_{ij} \)) and profits (\( p_i \)) are highly correlated are more difficult to solve. When high-profit items also have high weights, selecting profitable items quickly exhausts available capacity, complicating the optimization process.
    
    \item \textbf{Tightness Ratio (\(\alpha\))}: The tightness ratio is a key parameter that controls how constrained the problem is. It is defined as:
    \begin{equation}
        \alpha_j = \frac{c_j}{\sum_{i=1}^{n} w_{ij}}, \quad \forall j \in \{1, \dots, m\}.
    \end{equation}
    A tightness ratio close to \( 1 \) means the total weight is nearly equal to the knapsack's capacity, making the problem highly constrained and difficult to solve. Decreasing \(\alpha\) creates a looser problem, whereas increasing it results in a more restrictive and computationally challenging scenario.

    \item \textbf{Scaling the Number of Items (\(n\))}: Increasing \(n\) enlarges the search space exponentially, making it harder for exact algorithms to find optimal solutions efficiently.
\end{enumerate}

By carefully adjusting these parameters, researchers can generate MDKP instances with varying difficulty levels, enabling effective benchmarking of optimization algorithms and analysis of their performance under different conditions.

For further details on MDKP, see \cite{puchinger2010multidimensional}.

\subsubsection{Benchmark Datasets}
To evaluate solution methodologies for the MDKP, we employ the SAC-94 dataset from \cite{dataset}, which is derived from a variety of real-world problems. It provides challenging instances of MDKP with varying numbers of items, constraints, and complexity, and also serves as a robust benchmark for testing both classical and advanced optimization techniques, offering diverse challenges and scalability.

\subsection{Maximum Independent Set}

The \textit{Maximum Independent Set} (MIS) problem \cite{mis} is a cornerstone problem in graph theory and combinatorial optimization. Given a graph \( G = (V, E) \), where \( V \) is the set of vertices and \( E \) is the set of edges, the objective is to find the largest subset \( I \subseteq V \) such that no two vertices in \( I \) are adjacent:
\begin{equation}
    \forall u, v \in I, \quad (u, v) \notin E.
\end{equation}

This problem arises in numerous domains, including wireless communication, task scheduling, and resource allocation. However, solving the MIS problem is computationally challenging, as it is classified as NP-hard. For large-scale graphs, direct solutions often become infeasible, necessitating preprocessing methods to reduce graph size and complexity.

\subsubsection{Mathematical Formulation}
The MIS problem can be formulated as a binary integer programming problem. Let \( x_i \) be a binary variable for each vertex \( i \in V \), where:
\begin{equation}
    x_i =
    \begin{cases}
        1 & \text{if vertex } i \text{ is in the independent set}, \\
        0 & \text{otherwise}.
    \end{cases}
\end{equation}

The objective is to maximize the size of the independent set:
\begin{equation}
    \max \sum_{i \in V} x_i,
\end{equation}
subject to the constraints:
\begin{equation}
    x_u + x_v \leq 1, \quad \forall (u, v) \in E.
\end{equation}

The constraints ensure that no two adjacent vertices are included in the independent set.

\subsubsection{Preprocessing Using Simplicial Nodes}

We make use of a polytime pre-processing technique for MIS \cite{kroger2024polytime}, which works on a recursive fixing procedure that generalizes the existing polytime algorithm to solve the maximum independent set problem on chordal graphs, which admit simplicial orderings.
A vertex \( v \in V \) is termed \textit{simplicial} if its neighborhood forms a clique:
\begin{equation}
    N(v) = \{u \in V \mid (v, u) \in E\}.
\end{equation}
The subgraph induced by \( N(v) \) must satisfy:
\begin{equation}
    \forall u, w \in N(v), \quad (u, w) \in E.
\end{equation}
Simplicial nodes are significant because they can be directly added to the independent set without ambiguity, simplifying the graph.

\subsubsection{Algorithm}

\begin{algorithm}[H]
\caption{Graph Preprocessing for Maximum Independent Set Problem}
\label{alg:preprocessing}

\textbf{Input:} Graph $G = (V, E)$ \\
\textbf{Output:} Reduced graph $G' = (V', E')$ and independent set $I$

\begin{algorithmic}[1]
\Procedure{PreprocessGraph}{$G$}
    \State Initialize $I \gets \emptyset$ (independent set)
    \While{there exist simplicial nodes in $G$}
        \State Identify all simplicial nodes $S \subseteq V$
        \State Add simplicial nodes to the independent set: $I \gets I \cup S$
        \State Remove simplicial nodes $S$ and their neighbors $N(S)$ from $G$:
        \[
        V \gets V \setminus (S \cup N(S)), \quad 
        E \gets E \setminus \{(u, v) \mid u, v \in S \cup N(S)\}
        \]
    \EndWhile
    \State \textbf{return} $G' = (V, E)$, $I$
\EndProcedure
\end{algorithmic}
\end{algorithm}

The reduced graph, now free of simplicial nodes, is solved using standard Maximum Independent Set (MIS) methods. The final independent set is then reconstructed by integrating the solution of the reduced graph with the simplicial nodes preserved during preprocessing.

The preprocessing step streamlines the problem by reducing the graph’s size and complexity. By removing nodes and edges associated with simplicial nodes, it eliminates trivial substructures, yielding a smaller graph that allows solvers to focus on the more computationally challenging regions.

The following Figure~\ref{fig:graph_reduction} demonstrates the iterative preprocessing applied to a graph instance. Nodes are color-coded as follows:
\begin{itemize}
    \item \textbf{Red Nodes}: Simplicial nodes added to the independent set.
    \item \textbf{Yellow Nodes}: Neighbors of simplicial nodes, fixed to zero.
    \item \textbf{Blue Nodes}: Remaining nodes in the graph.
\end{itemize}

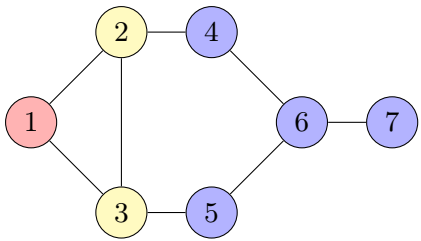
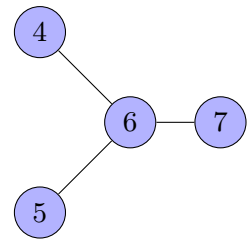
\begin{figure*}[ht!]
    \centering
    \begin{subfigure}[b]{0.45\textwidth}
        \centering
        \begin{tikzpicture}[scale=1.2, transform shape]
            \node[circle, draw, fill=red!30] (A) at (0, 0) {1};
            \node[circle, draw, fill=yellow!30] (B) at (1, 1) {2};
            \node[circle, draw, fill=yellow!30] (C) at (1, -1) {3};
            \node[circle, draw, fill=blue!30] (D) at (2, 1) {4};
            \node[circle, draw, fill=blue!30] (E) at (2, -1) {5};
            \node[circle, draw, fill=blue!30] (F) at (3, 0) {6};
            \node[circle, draw, fill=blue!30] (G) at (4, 0) {7};
            \draw (A) -- (B);
            \draw (A) -- (C);
            \draw (B) -- (C);
            \draw (B) -- (D);
            \draw (C) -- (E);
            \draw (D) -- (F);
            \draw (E) -- (F);
            \draw (F) -- (G);
        \end{tikzpicture}
        \caption{Iteration 1: Identify and remove simplicial nodes (red).}
        \label{fig:iteration1}
    \end{subfigure}
    \hfill
    \begin{subfigure}[b]{0.45\textwidth}
        \centering
        \begin{tikzpicture}[scale=1.2, transform shape]
            \node[circle, draw, fill=blue!30] (D) at (2, 1) {4};
            \node[circle, draw, fill=blue!30] (E) at (2, -1) {5};
            \node[circle, draw, fill=blue!30] (F) at (3, 0) {6};
            \node[circle, draw, fill=blue!30] (G) at (4, 0) {7};
            \draw (D) -- (F);
            \draw (E) -- (F);
            \draw (F) -- (G);
        \end{tikzpicture}
        \caption{Iteration 2: Reduced graph after removing neighbors.}
        \label{fig:iteration2}
    \end{subfigure}
    \caption{Graph reduction through preprocessing. In each iteration, simplicial nodes (red) and their neighbors (yellow) are removed, leaving a progressively reduced graph. Blue nodes represent the remaining graph nodes.}
    \label{fig:graph_reduction}
\end{figure*}

\subsubsection{Complexity Factors and Parameter Adjustments in the Maximum Independent Set Problem}

The Maximum Independent Set (MIS) problem is an NP-hard combinatorial optimization problem, with its computational complexity influenced by several factors. Key determinants of problem difficulty include:

\begin{itemize}
    \item \textbf{Graph Density and Vertex Degree:} Higher graph density (ratio of existing edges to possible edges) reduces the number of non-adjacent vertices available for an independent set, making the problem harder. Similarly, graphs with high-degree vertices limit feasible selections, as choosing one excludes many neighboring vertices.
    
    \item \textbf{Graph Structure and Planarity:} Graphs with a high chromatic number (requiring more colors for proper vertex coloring) tend to have intricate structures, making large independent sets harder to identify. Non-planar graphs and irregular topologies further increase computational complexity.
\end{itemize}

To generate harder MIS instances for benchmarking, one can:
\begin{itemize}
    \item \textbf{Increase Graph Density and Vertex Degree:} Densifying the graph and incorporating high-degree vertices constrain the search space, reducing the number of valid independent sets.
    \item \textbf{Use Complex Graph Topologies:} Employing non-planar graphs or those with high chromatic numbers removes structural simplifications, increasing computational difficulty.
\end{itemize}

By tuning these parameters, researchers can modulate the difficulty of MIS instances, enabling more effective benchmarking of optimization algorithms across different problem complexities.

\subsubsection{Benchmark Datasets}

To evaluate our approach, we use well-established benchmark datasets derived from graphs representing error-correcting codes \cite{sloane2000challenge}. These graphs, known for their combinatorial complexity, provide challenging instances for the Maximum Independent Set (MIS) problem. The datasets are described below:

\paragraph{Single-Deletion-Correcting Codes.} 
We use the graph \textbf{1dc.64.txt}, a 64-node graph where the known size of its maximal independent set is $10$, representing the codewords capable of correcting a single deletion.

\paragraph{Single Transposition-Correcting Codes (Excluding End-Around Transpositions).} 
These graphs decompose into \(k+1\) connected components, where \(k = \log_2(\text{number of nodes})\). We use the following instances:
\begin{itemize}
    \item \textbf{1tc.8.txt}: 8 nodes, MIS size = 4.
    \item \textbf{1tc.16.txt}: 16 nodes, MIS size = 8.
    \item \textbf{1tc.32.txt}: 32 nodes, MIS size = 12.
    \item \textbf{1tc.64.txt}: 64 nodes, MIS size = 20.
\end{itemize}

\paragraph{Single Transposition-Correcting Codes (Including End-Around Transpositions).} 
We include \textbf{1et.64.txt}, a 64-node graph with a maximal independent set size of 10. The inclusion of end-around transpositions introduces additional structural complexity.

These graphs provide diverse and scalable challenges due to their origins in error-correcting codes. The known sizes of maximal independent sets offer a reliable ground truth for assessing solution accuracy, making them ideal benchmarks for comparing classical and advanced optimization techniques.

\subsection{Quadratic Assignment Problem}

The \textit{Quadratic Assignment Problem} (QAP)\cite{qap} 
was first introduced by Koopmans and Beckmann in 1957 \cite{qap} to model the optimal assignment of economic activities. Since then, it has found applications in various fields. It is a fundamental combinatorial optimization problem that models the assignment of a set of facilities to a set of locations, aiming to minimize the total cost associated with the assignment. Each pair of facilities has a flow between them, and each pair of locations has a distance; the objective is to assign facilities to locations such that the sum of the products of flows and corresponding distances is minimized.

\subsubsection{Mathematical Formulation}

Formally, given \( n \) facilities and \( n \) locations, let:

\begin{itemize}
    \item \( F = [f_{ij}] \) be an \( n \times n \) flow matrix, where \( f_{ij} \) represents the flow between facilities \( i \) and \( j \).
    \item \( D = [d_{kl}] \) be an \( n \times n \) distance matrix, where \( d_{kl} \) denotes the distance between locations \( k \) and \( l \).
\end{itemize}

The goal is to find a permutation \( \pi \) of \( \{1, 2, \ldots, n\} \) that minimizes the objective function:

\[
\text{Minimize} \quad \sum_{i=1}^{n} \sum_{j=1}^{n} f_{ij} \, d_{\pi(i)\pi(j)}
\]

Alternatively, this can be expressed using permutation matrices. Let \( P \) be an \( n \times n \) permutation matrix corresponding to the permutation \( \pi \). The objective function can then be written as:

\[
\text{Minimize} \quad \text{trace}(F P D P^T)
\]

where \( \text{trace}(\cdot) \) denotes the trace of a matrix, which is the sum of its diagonal elements. The permutation matrix \( P \) satisfies the following constraints:

\[
P \mathbf{e} = \mathbf{e}, \quad P^T \mathbf{e} = \mathbf{e}, \quad P_{ij} \in \{0, 1\} \quad \forall i, j
\]

Here, \( \mathbf{e} \) is a column vector of ones of appropriate dimension. The first two constraints ensure that \( P \) is a doubly stochastic matrix (each row and each column sums to one), and the third constraint enforces that \( P \) is a permutation matrix.

\subsubsection{Factors Affecting Problem Complexity and Hardness}

The QAP is a fundamental combinatorial optimization problem recognized for its significant computational complexity. In general, instances of size $n > 30$ cannot be solved in reasonable time. Several factors contribute to the difficulty of solving QAP, and certain parameter adjustments can further increase its complexity:

\begin{enumerate}
    \item \textbf{Problem Size (Number of Facilities/Locations):} As the number of facilities and locations (\(n\)) increases, the solution space grows factorially (\(n!\)), making exhaustive search methods computationally infeasible for large \(n\). Even for moderate values of \(n\), finding the optimal assignment becomes challenging due to the vast number of possible permutations \cite{qaplib}.

    \item \textbf{Flow and Distance Matrix Characteristics:}
    \begin{itemize}
        \item \textbf{Matrix Sparsity:} Sparse flow or distance matrices, where many entries are zero, can simplify the problem since fewer facility-location interactions need to be considered. Conversely, dense matrices increase complexity due to the multitude of non-zero interactions.
        \item \textbf{Correlation Between Matrices:} The relationship between the flow and distance matrices affects problem difficulty. Instances where high-flow pairs correspond to long distances can lead to higher costs, complicating the optimization process.
    \end{itemize}

    \item \textbf{Symmetry in the Assignment:} Symmetric problems, where multiple assignments yield the same cost, can complicate the search for unique optimal solutions. This symmetry can cause algorithms to explore redundant solutions, increasing computational effort.
\end{enumerate}

To increase the difficulty of QAP instances, one can adjust the following parameters:

\begin{itemize}
    \item \textbf{Increasing Problem Size:} Expanding the number of facilities and locations (\(n\)) exponentially increases the number of possible assignments, thereby escalating computational complexity.

    \item \textbf{Enhancing Matrix Density:} Populating the flow and distance matrices with more non-zero values (increasing density) introduces additional interactions between facilities and locations, leading to a more complex cost landscape.

    \item \textbf{Introducing Asymmetry:} Designing problems where the flow and/or distance matrices are asymmetric removes potential simplifications from symmetric properties, making the problem more challenging to solve.
\end{itemize}

\subsubsection{Benchmark Datasets}

To facilitate research and benchmarking of solution methods, \textit{QAPLIB} \cite{qaplib} was established as a comprehensive repository of Quadratic Assignment Problem (QAP) instances and solutions. It serves as a standard testbed, offering problem instances of varying sizes and complexities, along with known optimal or best-known solutions. QAPLIB is a crucial resource for evaluating algorithmic performance and benchmarking new approaches against established solutions.

In this study, several benchmark instances from QAPLIB were utilized to assess the effectiveness of the proposed solution approach. These instances, spanning different problem sizes and complexities, provided a rigorous framework for testing and validation. By comparing the results of the proposed method with known QAPLIB solutions, its efficiency and accuracy were systematically evaluated.

\subsection{Market Share Problem}

The Market Share Problem \cite{market_share} is a combinatorial optimization problem that models the allocation of products to retailers while minimizing deviations from a desired market split. 

\subsubsection{Mathematical Formulation}

This problem can be mathematically formulated as:

\begin{equation}
\text{Minimize } \sum_{i=1}^m |s_i|
\end{equation}
subject to:
\begin{equation}
\sum_{j=1}^n a_{ij}x_j + s_i = b_i, \quad i = 1, \dots, m,
\end{equation}
\begin{equation}
x_j \in \{0, 1\}, \quad j = 1, \dots, n,
\end{equation}
\begin{equation}
s_i \text{ is free, for } i = 1, \dots, m.
\end{equation}

Here:
\begin{itemize}
    \item \(n\) is the number of products,
    \item \(m\) is the number of retailers,
    \item \(a_{ij}\) is the demand of retailer \(i\) for product \(j\),
    \item \(b_i\) is the desired market share target for retailer \(i\),
    \item \(x_j \in \{0, 1\}\) indicates whether product \(j\) is assigned (\(x_j = 1\)) or not (\(x_j = 0\)),
    \item \(s_i\) is a slack variable to account for deviations from the desired allocation.
\end{itemize}

The objective function minimizes the total deviation from the desired allocation of products, as captured by the slack variables \(s_i\). The constraints enforce that the total demand met for each retailer \(i\) is either exactly or approximately equal to the target \(b_i\). 

This problem also corresponds to a \textbf{Feasibility Problem (FP)} in geometry:
\begin{quote}
\textit{Given \(m\) hyperplanes in \(\mathbb{R}^n\), does there exist a point \(x \in \{0,1\}^n\) that lies on the intersection of these \(m\) hyperplanes?}
\end{quote}

If the optimal value of the objective function is \(0\), the answer is "yes," indicating the allocation perfectly satisfies the targets. If not, the answer is "no." For \(m = 1\), this problem reduces to the well-known \textbf{subset-sum problem}, which is NP-complete.

\subsection*{\texorpdfstring{Determining the Range of \(s_i\)}{Determining the Range of sᵢ}}

From the equality constraint:
\begin{equation}
s_i = b_i - \sum_{j=1}^n a_{ij} x_j,
\end{equation}
the range of \(s_i\) depends on the extreme values of the term \(\sum_{j=1}^n a_{ij}x_j\):
\begin{itemize}
    \item When all \(x_j = 0\), the sum becomes \(0\),
    \item When all \(x_j = 1\), the sum becomes \(\sum_{j=1}^n a_{ij}\).
\end{itemize}

Hence, \(s_i\) can take values in the range:
\begin{equation}
b_i - \sum_{j=1}^n a_{ij} \leq s_i \leq b_i.
\end{equation}
The absolute maximum deviation of \(s_i\) occurs when \(x_j\) minimizes or maximizes the left-hand side:
\begin{equation}
\max\{|b_i|, |b_i - \sum_{j=1}^n a_{ij}|\}.
\end{equation}

\subsubsection{Generating Hard Instances of the Market Share Problem}

To create challenging instances of the Market Share Problem, specific configurations of parameters \(a_{ij}\), \(b_i\), \(n\), and \(m\) are chosen. These configurations are designed to produce problem instances that are computationally difficult for traditional integer programming solvers.

The parameters for generating hard instances are defined as follows:
\begin{itemize}
    \item \textbf{Demand Matrix (\(a_{ij}\)):} The entries \(a_{ij}\) represent the demand of retailer \(i\) for product \(j\). These are sampled as uniform random integers in the range:
    \[
    a_{ij} \in [0, 99].
    \]
    This range ensures a diverse and challenging demand structure across retailers.

    \item \textbf{Market Share Targets (\(b_i\)):} The target market shares \(b_i\) are computed based on the desired split of products. For a 50/50 split between two divisions \(D_1\) and \(D_2\), \(b_i\) is set as:
    \[
    b_i = \frac{1}{2} \sum_{j=1}^n a_{ij}.
    \]
    More generally, \(b_i\) can be chosen in the range:
    \[
    b_i \in \left[\frac{1}{2} \left(-D + \sum_{j=1}^n a_{ij}\right), \frac{1}{2} \left(-D + \sum_{j=1}^n a_{ij}\right) + D - 1\right],
    \]
    where \(D\) is a parameter that controls the variability of \(b_i\).

    \item \textbf{Number of Retailers (\(m\)) and Products (\(n\)):} To generate challenging instances, the number of products \(n\) is set proportional to the number of retailers \(m\). A recommended configuration is:
    \[
    n = 10(m - 1),
    \]
    which ensures that the problem grows in complexity as \(m\) increases.
\end{itemize}

\subsubsection{Example of Hard Instance Generation}
Consider the following configuration:
\begin{itemize}
    \item \(D = 100\),
    \item \(m = 10\),
    \item \(n = 10(m - 1) = 90\).
\end{itemize}

The steps to generate the instance are:
\begin{enumerate}
    \item Generate the demand matrix \(a_{ij}\) with each \(a_{ij}\) sampled uniformly at random from \([0, 99]\).
    \item Compute \(b_i\) for each retailer \(i\) as:
    \[
    b_i = \frac{1}{2} \sum_{j=1}^n a_{ij}.
    \]
    \item Formulate the problem by assigning \(x_j \in \{0, 1\}\) to represent product allocation.
\end{enumerate}

This configuration creates a class of hard instances, where the solver must balance the product allocation across \(m\) retailers to meet the target market shares.

\subsubsection{Key Challenges of Hard Instances}

\begin{itemize}
    \item \textbf{High Dimensionality:} For large \(m\) and \(n\), the search space grows exponentially, making it challenging to explore all possible allocations.
    \item \textbf{Random Demand Values:} The randomness in \(a_{ij}\) introduces variability, increasing the difficulty of finding optimal solutions.
    \item \textbf{Feasibility Complexity:} Even determining whether a feasible solution exists (where all slack variables \(s_i = 0\)) is an NP-complete problem.
\end{itemize}

\subsection*{\texorpdfstring{QUBO Formulation}{Rationale for Splitting |s| into s+ and s-}}

The original formulation contains absolute value function \(|s|\) which is non-linear and non-differentiable at \(s = 0\), making it incompatible with standard optimization frameworks.

We apply a standard trick to split \(|s|\) into \(s^+\) and \(s^-\) to represent the function in a linear form:
\begin{equation}
|s| = s^+ + s^- \quad \text{and} \quad s = s^+ - s^-,
\end{equation}
where \(s^+ \geq 0\) and \(s^- \geq 0\).

Hence, the revised objective becomes:
\begin{equation}
\text{Minimize } \sum_{i=1}^m (s_i^+ + s_i^-).
\end{equation}

And similarly, the revised constraints become: 

\begin{equation}
\sum_{j=1}^n a_{ij} x_j + (s_i^+ - s_i^-) = b_i.
\end{equation}

To represent \(s^+\) and \(s^-\) in QUBO, use binary encoding:
\begin{equation}
s_i^+ = \sum_{k=0}^{\log_2(U)} 2^k z_{i,k}^+, \quad s_i^- = \sum_{k=0}^{\log_2(U)} 2^k z_{i,k}^-,
\end{equation}
where \(z_{i,k}^+, z_{i,k}^- \in \{0, 1\}\), and \(U\) is the upper bound for \(s^+\) and \(s^-\).

\textbf{Example:}

If \(b_i = 50\) and \(\sum_{j=1}^n a_{ij} = 100\), the upper bound for \(s^+\) and \(s^-\) is:
\begin{equation}
\max\{|b_i|, |b_i - \sum_{j=1}^n a_{ij}|\} = \max\{50, 50\} = 50.
\end{equation}
Thus, \(s^+\) and \(s^-\) require:
\begin{equation}
\lceil \log_2(50 + 1) \rceil = 6 \text{ binary variables each}.
\end{equation}

\subsubsection{Benchmark Datasets}

Following the guidelines of \cite{market_share}, we generate hard problem instances with the number of retailers ranging from $3$ to $10$, while the remaining parameters are set based on the discussion above. These instances are then solved using CPLEX as the classical baseline, and the results are presented in Table \ref{tab:cplex_performance_metrics}. 

The code for generating test instances and solving them with CPLEX is available on GitHub \cite{Sharma2025QOB}. Details about the performance of CPLEX on these hard instances is given in the Appendix~ \ref{app:performance}.

\twocolumngrid 
\section{\label{sec:level7}Experimental Setup}
\subsection{Environment}

The experiments were conducted on a high-performance server equipped with an Intel(R) Xeon(R) Gold 6154 CPU @ 3.00GHz, featuring 144 CPUs across four sockets, with 18 cores per socket and two threads per core. Quantum computations were simulated using the \texttt{Qiskit AerSimulator} \cite{qiskit2024} with the matrix product state method.

\subsection{Algorithm Details}

For reproducibility, we provide detailed specifications for each quantum algorithm used in our experiments. For QAOA and its variants, we used the default QAOA ansatz, with depth and gate parameters detailed in the Appendix~\ref{app:metrics}. For VQE, CVaR VQE, and QRAO, we employed an Efficient SU2 ansatz with varying depths depending on the problem instance (specific values are provided in Appendix~\ref{app:metrics}). For CVaR (Expected Shortfall) variants, the confidence level $\alpha$ is defined within the range $(0,1]$. In this study, we set $\alpha=0.25$. Notably, as $\alpha \rightarrow 0$, CVaR converges to the minimum value, whereas $\alpha = 1$ corresponds to the expected value of the random variable \cite{Barkoutsos_2020}.

In this work, we utilize Pauli Correlation Encoding (PCE) with a QUBO-based cost function and a dynamic multi-step re-optimization scheme to efficiently solve combinatorial optimization problems on constrained quantum hardware. Specifically, we replace the traditional Weighted Max-Cut loss with a more flexible QUBO formulation and introduce controlled perturbations alongside exhaustive local searches to escape local minima and improve solution quality. To parameterize the quantum state, we employ a Brickwork ansatz consisting of single-qubit rotations and entangling $R_{XX}$ layers. Our optimization procedure adaptively balances exploration and exploitation through dynamic perturbation scaling and iterative refinement, enhancing convergence. Full algorithmic details, including the mathematical formulation, circuit design, and re-optimization strategy, are provided in Appendix~\ref{app:pce}.

Classical optimizers such as POWELL and COBYLA were applied with a maximum of 5000 function evaluations and iterations. For PCE, the SLSQP optimizer was used with a maximum of 100 iterations. Notably, for PCE, only a single-step optimization was performed, despite the implementation supporting multiple re-optimization steps. This choice was made to ensure fairness across all implementations, as we did not employ any recursive algorithms, and to mitigate excessive computational time constraints. Additionally, a multi-bit swap operation was employed as a classical post-processing step.

Across all algorithms, parameters were randomly initialized within the range 
\(-\pi\) to \(\pi\) , and each circuit execution was performed with 4000 shots, balancing computational cost with accuracy, as higher shot counts significantly increase resource requirements and runtime. In qubit-efficient techniques like QRAO, the compression parameter was set to 3 to achieve the highest possible compression while maintaining feasible circuit depths, resulting in an average compression rate of ~70–80\%. For PCE, the number of Pauli correlations was fixed at 2 to enforce quadratic compression, ensuring a reasonable trade-off between qubit savings and circuit depth. While higher compression levels are possible, they require significantly deeper circuits, making them impractical for our setup. Detailed information on the number of qubits retained after compression is provided in the Appendix~\ref{app:metrics}.

All experiments were simulated using Qiskit, utilizing the \texttt{backend sampler v2} and \texttt{backend estimator v2} with the matrix product state method.

\subsection{Metrics}

The performance of the proposed approach was evaluated using the following metrics:
\begin{itemize}
    \item \textbf{Optimality Gap (\%)}: The percentage difference between the obtained solution and the known optimal solution.
\begin{align}
\text{Opt. Gap (\%)} = 
    & \frac{
    \text{Obj. Best} 
    - \text{Obj. Obtained}
    }{
    \text{Obj. Best}
    } \nonumber 
    & \times 100
\end{align}

    \item \textbf{Relative Solution Quality (\%):} This metric evaluates the quality of the obtained solution relative to the best-known or optimal solution. It is typically expressed as a percentage, calculated as:
\begin{equation}
\text{RSQ (\%)} = \left( 
    \frac{\text{Obj. Value of Obtained Solution}}
         {\text{Obj. Value of Best-Known Solution}} 
\right) \times 100
\end{equation}
Higher values indicate solutions closer to the optimal, demonstrating the effectiveness of the algorithm or approach used.

\end{itemize}

Table~\ref{tab:results_mdkp} presents the results of our experiments for MDKP, while Tables~\ref{tab:mis_pce_qrao_results}, \ref{tab:mis_vqe_results}, and \ref{tab:mis_qaoa_results} summarize the results for MIS problem instances using various quantum methods. Similarly, Table~\ref{tab:results_qap} presents the QAP results, and Table~\ref{tab:results_ms} presents the MSP results. More details on each can be found in their respective sections.

\paragraph{Quantum Resource Usage:} 
In addition to solution quality metrics, we provide a detailed breakdown of the quantum resources required (see Appendix~\ref{app:metrics}) for each instance and method. This includes the number of qubits, ansatz depth, total gate count, number of two-qubit gates, trainable parameters, and execution time (in minutes). These resource metrics offer valuable insights into the feasibility and scalability of different quantum optimization techniques across various problem instances.

\section{\label{sec:level8}Results and Analysis}

This section presents the experimental results and performance evaluation of the proposed approach. We analyze solution quality, computational efficiency, and resource utilization across various problem instances. Comparisons with classical and quantum baselines highlight the strengths and limitations of different methods. Additionally, we examine the scalability of the proposed techniques and discuss key insights derived from the results.

In our benchmarking experiments, we observed significant differences in runtime among the evaluated quantum algorithms. In particular,QAOA, its variants, and QRAO exhibited notably slower performance on dense instances of MDKP, QAP, and MSP, where high density led to a substantial increase in two-qubit gates, significantly raising the computational cost of simulations. For instance, the densest VQE circuit with comparable depth completed execution in approximately 700 minutes (11 hours), whereas QAOA failed to complete even a single full optimization cycle within 24 hours. This discrepancy is partly due to the fact that while VQE’s entanglement was constrained to a linear pattern, QAOA’s ansatz inherently introduces more complex, uncontrolled entanglement, further compounding runtime challenges.

By contrast, the Maximum Independent Set (MIS) instances we considered were less dense. While a 64-node instance with high edge density resulted in a memory error, the instance that successfully ran was significantly sparser. Additionally, the Variational Quantum Eigensolver (VQE) and its variants demonstrated more efficient execution, primarily due to the flexibility of their customizable ansatz—a feature not shared by QAOA. This tunability allowed VQE to handle dense problems more effectively. However, for the densest problem instances, even QRAO was unable to achieve a compression ratio greater than 2.

\subsection{MDKP Results}

To evaluate the effectiveness of quantum optimization methods for MDKP, we compare the performance of \textbf{PCE}, \textbf{VQE}, and \textbf{CVaR VQE} on benchmark instances. Our analysis, summarized in Table~\ref{tab:results_mdkp}, focuses on three critical aspects: \textit{solution feasibility, optimality gap, and qubit efficiency}.

\paragraph{Feasibility Across Methods}
The ability to find feasible solutions is crucial for practical applications. Our results show that:
\begin{itemize}
    \item PCE produces feasible solutions across all instances, demonstrating the robustness of its encoding strategy when paired with classical post-processing.
    \item VQE fails to find a feasible solution for instance pb4, highlighting its sensitivity to parameter initialization and optimization.
    \item CVaR VQE maintains feasibility across all instances, reinforcing the benefits of minimizing CVaR over standard expectation-based methods.
    \item Notably, classical solvers like CPLEX easily find optimal solutions for these small instances, highlighting the current gap between quantum and classical methods. While CVaR VQE improves feasibility, all quantum approaches still exhibit notable optimality gaps compared to classical solutions.
\end{itemize}

\paragraph{Optimality Gap and Solution Quality}
The optimality gap is a key indicator of solution quality, with lower values indicating better performance. Our results indicate:
\begin{itemize}
    \item PCE achieves a lower optimality gap than VQE in most instances, suggesting its encoding effectively captures problem constraints.
    \item CVaR VQE does not consistently outperform PCE, with instances such as pet5, pet6, and pet7 showing higher gaps than both PCE and standard VQE.
    \item VQE performs worst in pb4, where it fails to find a feasible solution, reinforcing the challenges of variational quantum algorithms in constrained combinatorial optimization.
\end{itemize}

\paragraph{Qubit Efficiency and Scalability}
Quantum resource efficiency is a major concern in near-term quantum devices. Our results highlight:
\begin{itemize}
    \item PCE significantly reduces the number of qubits required, utilizing only 6 to 10 qubits per instance, making it a more hardware-efficient encoding.
    \item The number of qubits in PCE scales sublinearly with problem size, whereas VQE and CVaR VQE require full QUBO encoding, leading to higher qubit counts in larger instances.
\end{itemize}

\paragraph{Key Takeaways}
These findings provide key insights into the potential of quantum optimization for MDKP:
\begin{itemize}
    \item PCE balances feasibility, solution quality, and qubit efficiency, making it a strong candidate for solving MDKP on near-term quantum devices.
    \item VQE struggles with feasibility and solution quality in constrained instances, suggesting a need for improved parameter optimization techniques.
    \item CVaR VQE uses CVaR as an aggregation function, but does not consistently improve optimality over standard VQE.
    \item Qubit-efficient encodings like PCE are crucial for scaling quantum optimization methods to larger problem instances.
\end{itemize}

\begin{table*}[t]
    \caption{\label{tab:results_mdkp}Performance comparison of Pauli Correlation Encoding (PCE), Variational Quantum Eigensolver (VQE), and Conditional Value-at-Risk VQE (CVaR VQE) on classically benchmarked hard instances of the Multi-Dimensional Knapsack Problem (MDKP). The "Variables" column indicates the number of variables in the quadratic unconstrained binary optimization (QUBO) formulation. Each method reports the number of qubits used (PCE only), feasibility (Feas: Yes/No), and the optimality gap (\%) relative to the known optimal solution.}
    \begin{ruledtabular}
    \begin{tabular*}{\textwidth}{@{\extracolsep{\fill}}c|c|c|ccc|cc|cc}
        \textbf{Instance} & \textbf{Optimal (Known)} & \textbf{Variables} & 
        \multicolumn{3}{c|}{\textbf{PCE}} & 
        \multicolumn{2}{c|}{\textbf{VQE}} & 
        \multicolumn{2}{c}{\textbf{CVaR VQE}} \\ 
        \cline{4-10}
        & & & \textbf{Qubits} & \textbf{Feas} & \textbf{Gap (\%)} & 
        \textbf{Feas} & \textbf{Gap (\%)} & 
        \textbf{Feas} & \textbf{Gap (\%)} \\ 
        \hline
        hp1 & 3418 & 60 & 7  & Yes & 20.88 & Yes & 39.76 & Yes & 20.59 \\
        hp2 & 3186 & 67 & 8  & Yes & 38.38 & Yes & 12.34 & Yes  & 21.78   \\
        pb1 & 3090 & 59 & 7  & Yes & 14.04 & Yes & 19.94 & Yes  & 23.43   \\
        pb2 & 3186 & 66 & 8  & Yes & 14.37 & Yes & 19.49 & Yes  & 22.78   \\
        pb4 & 95168 & 45 & 6 & Yes & 32.91 & No  & inf.  & Yes  & 39.38   \\
        pb5 & 2139 & 116 & 10 & Yes & 11.87 & Yes & 4.25 & Yes  & 33.34   \\
        pet2 & 87061 & 99 & 9 & Yes & 28.19 & Yes & 41.07 & Yes  & 30.37   \\
        pet3 & 4015 & 102 & 9 & Yes & 15.56 & Yes & 4.98 & Yes  & 34.74   \\
        pet4 & 6120 & 107 & 9 & Yes & 50.98 & Yes & 66.58 & Yes  & 48.44   \\
        pet5 & 12400 & 122 & 10 & Yes & 22.74 & Yes & 33.23 & Yes  & 53.15   \\
        pet6 & 10618 & 86 & 9 & Yes & 33.84 & Yes & 12.50 & Yes  & 40.41   \\
        pet7 & 16537 & 100 & 9 & Yes & 15.87 & Yes & 43.46 & Yes  & 46.47   \\
    \end{tabular*}
    \end{ruledtabular}
\end{table*}

\begin{figure}[h]
    \centering
    \includegraphics[width=\linewidth]{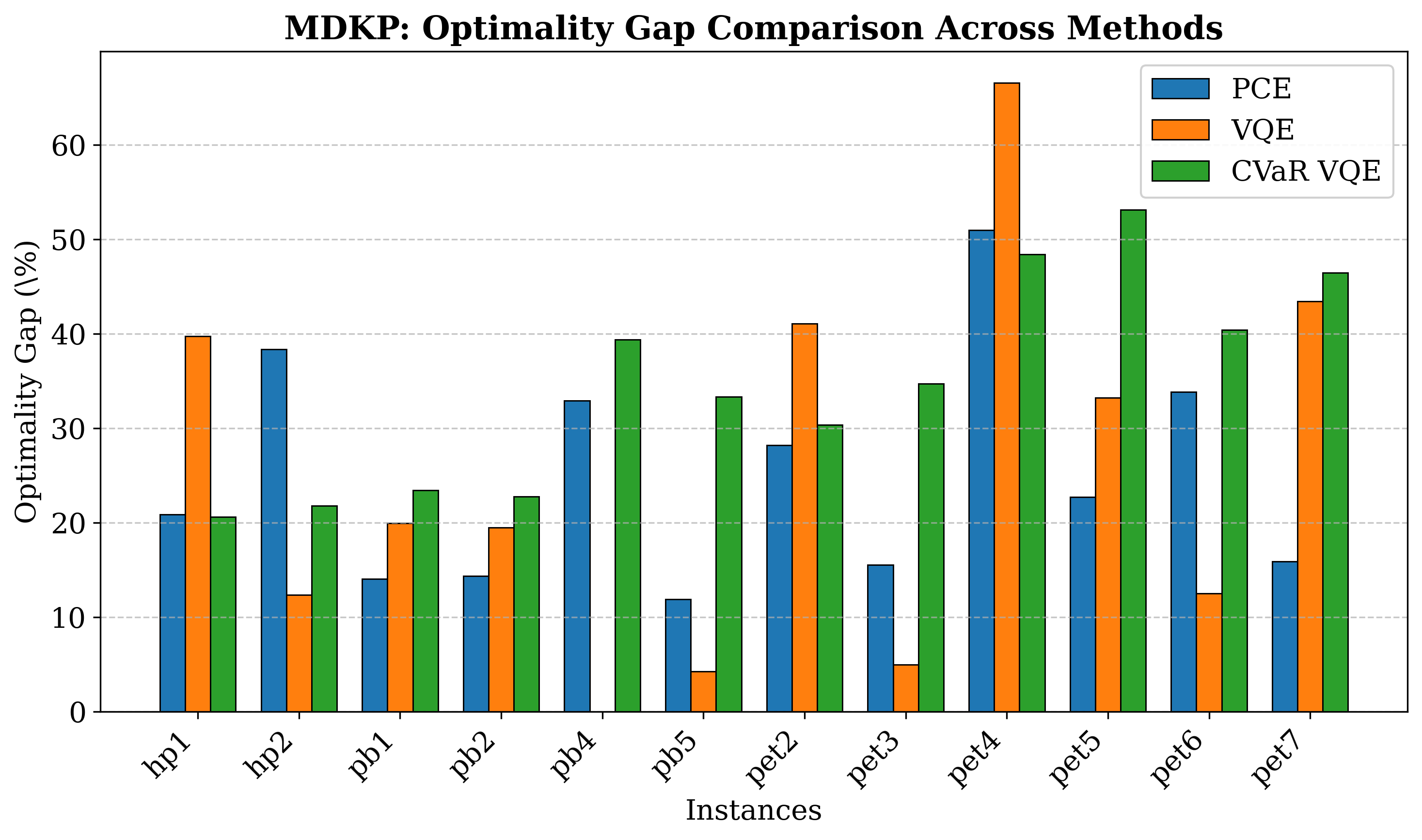}
    \caption{Comparison of optimality gaps for PCE, VQE, and CVaR VQE across MDKP instances. Lower gaps indicate better solution quality. PCE generally achieves lower optimality gaps compared to VQE, while CVaR VQE does not consistently outperform PCE. The y-axis is dynamically scaled to ensure all data points are visible.}

    \label{fig:mdkp_opt_gap}
\end{figure}

\subsection{MIS Results}

The performance of \textbf{PCE}, \textbf{QRAO}, \textbf{VQE}, \textbf{CVaR VQE}, and \textbf{QAOA and its variants} on benchmark Maximum Independent Set (MIS) instances is summarized in Tables~\ref{tab:mis_pce_qrao_results}, \ref{tab:mis_vqe_results}, and \ref{tab:mis_qaoa_results}. The evaluation is based on three key aspects: \textit{solution feasibility, relative solution quality (RSQ\%), and qubit efficiency}. 

\paragraph{Comparison of Qubit-Efficient Methods: PCE vs. QRAO}
\begin{itemize}
    \item PCE consistently requires fewer qubits than QRAO while maintaining high relative solution quality (RSQ\%), making it a more hardware-efficient approach.
    \item PCE achieves higher RSQ\% than QRAO in most instances, except for 1tc.16, where QRAO performs slightly better.
    \item Both methods maintain feasibility across all instances, demonstrating their ability to generate valid independent sets.
\end{itemize}

\paragraph{Variational Methods: VQE and CVaR VQE}
\begin{itemize}
    \item VQE and CVaR VQE achieve 100\% RSQ in smaller instances (1tc.8) but show variability in larger graphs.
    \item CVaR VQE outperforms standard VQE in instances like 1tc.32 but performs worse in others (e.g., 1dc.64).
    \item Both approaches remain feasible across all instances, suggesting their robustness in producing valid independent sets.
\end{itemize}

\paragraph{QAOA and Its Variants}
\begin{itemize}
    \item Standard QAOA struggles with feasibility, failing on instances 1et.64 and 1dc.64.
    \item Multi-Angle QAOA (MA QAOA) exhibits infeasibility in more cases, highlighting the difficulty in parameter optimization.
    \item When feasible, QAOA and its variants do not consistently outperform VQE-based methods in RSQ\%.
\end{itemize}

\paragraph{Key Takeaways}
\begin{itemize}
    \item PCE is the most qubit-efficient method, achieving high RSQ\% with fewer qubits.
    \item VQE-based methods perform consistently well, maintaining feasibility and competitive RSQ\%.
    \item QAOA and MA QAOA struggle with feasibility, making them less reliable for solving MIS in their current form.
    \item Qubit-efficient encodings like PCE and QRAO remain crucial for scaling MIS solutions on near-term quantum devices.
\end{itemize}

\begin{table*}
    \caption{\label{tab:mis_pce_qrao_results}Performance comparison of Pauli Correlation Encoding (PCE) and Quantum Random Access Optimization (QRAO) on benchmark instances of the Maximum Independent Set (MIS) problem. "PCE (Qubits, Feas, RSQ\%)" and "QRAO (Qubits, Feas, RSQ\%)" indicate the number of qubits used, feasibility, and the Relative Solution Quality (RSQ).}
    \begin{ruledtabular}
    \begin{tabular*}{\textwidth}{@{\extracolsep{\fill}}c|c|c|ccc|ccc}
        \textbf{Instance} & \textbf{Optimal (Known)} & \textbf{Variables} & 
        \multicolumn{3}{c|}{\textbf{PCE}} & 
        \multicolumn{3}{c}{\textbf{QRAO}} \\ 
        \cline{4-9}
        & & & \textbf{Qubits} & \textbf{Feasible} & \textbf{RSQ\%} & \textbf{Qubits} & \textbf{Feasible} & \textbf{RSQ\%} \\ 
        \hline
        1dc.64  & 8  & 64  & 7 & Yes & 87.5 & 18 & Yes & 75.0 \\
        1tc.16  & 8  & 16  & 4 & Yes & 100.0 & 6 & Yes & 87.5 \\
        1tc.32  & 12 & 32  & 6 & Yes & 100 & 13 & Yes & 58.33 \\
        1et.64  & 18 & 64  & 7 & Yes & 88.9 & 24 & Yes & 72.22 \\
        1tc.64  & 20 & 64  & 8 & Yes & 95.0 & 23 & Yes & 40.0 \\
        1tc.8   & 4  & 8   & 3 & Yes & 100.0 & 4 & Yes & 100.0 \\
    \end{tabular*}
    \end{ruledtabular}
\end{table*}

\begin{table*}
    \caption{\label{tab:mis_vqe_results}Performance comparison of Variational Quantum Eigensolver (VQE) and Conditional Value-at-Risk VQE (CVaR VQE) on benchmark instances of the Maximum Independent Set (MIS) problem. "Feas" indicates feasibility of the solution, and "RSQ\%" represents the Relative Solution Quality.}
    \begin{ruledtabular}
    \begin{tabular*}{\textwidth}{@{\extracolsep{\fill}}c|c|c|cc|cc}
        \textbf{Instance} & \textbf{Optimal (Known)} & \textbf{Variables} & 
        \multicolumn{2}{c|}{\textbf{VQE}} & 
        \multicolumn{2}{c}{\textbf{CVaR VQE}} \\ 
        \cline{4-7}
        & & & \textbf{Feas} & \textbf{RSQ\%} & \textbf{Feas} & \textbf{RSQ\%} \\ 
        \hline
        1dc.64  & 8  & 64  & Yes & 87.5 & Yes & 62.5 \\
        1tc.16  & 8  & 16  & Yes & 75.0 & Yes & 87.5 \\
        1tc.32  & 12 & 32  & Yes & 75.0 & Yes & 91.7 \\
        1et.64  & 18 & 64  & Yes & 77.8 & Yes & 77.8 \\
        1tc.64  & 20 & 64  & Yes & 40.0 & Yes & 40.0 \\
        1tc.8   & 4  & 8   & Yes & 100.0 & Yes & 100.0 \\
    \end{tabular*}
    \end{ruledtabular}
\end{table*}

\begin{table*}
    \caption{\label{tab:mis_qaoa_results}Performance comparison of Quantum Approximate Optimization Algorithm (QAOA) and its variants (MA QAOA, CVaR QAOA) on benchmark instances of the Maximum Independent Set (MIS) problem. "Feas" indicates feasibility of the solution, and "RSQ\%" represents the Relative Solution Quality. "ME" refers to memory error.}
    \begin{ruledtabular}
    \begin{tabular*}{\textwidth}{@{\extracolsep{\fill}}c|c|c|cc|cc}
        \textbf{Instance} & \textbf{Optimal} & \textbf{Variables} & 
        \multicolumn{2}{c|}{\textbf{QAOA}} & 
        \multicolumn{2}{c}{\textbf{MA QAOA}} \\ 
        \cline{4-7}
        & & & \textbf{Feas} & \textbf{RSQ\%}  & \textbf{Feas} & \textbf{RSQ\%}  \\ 
        \hline
        1dc.64  & 8  & 64  & ME & ME &  No & inf \\
        1tc.16  & 8  & 16  & Yes & 100.0  & Yes & 50.0 \\
        1tc.32  & 12 & 32  & Yes & 83.3 & Yes & 91.7  \\
        1et.64  & 18 & 64  & No & inf & No & inf  \\
        1tc.64  & 20 & 64  & Yes & 50.0 & No & inf  \\
        1tc.8   & 4  & 8   & Yes & 100.0 & Yes & 100.0  \\
    \end{tabular*}
    \end{ruledtabular}
\end{table*}

\begin{figure}[t]
    \centering
    \includegraphics[width=\linewidth]{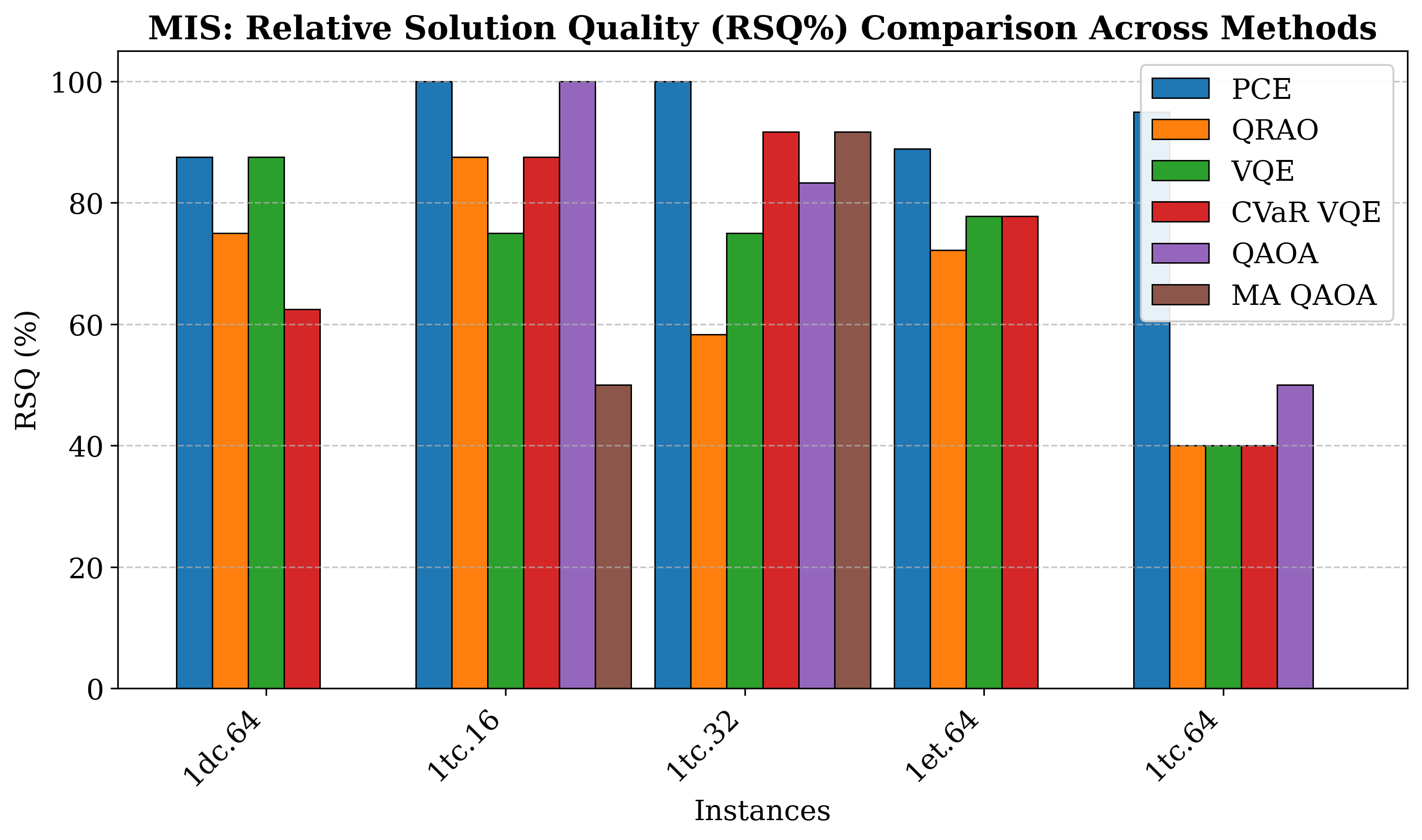}
    \caption{Comparison of Relative Solution Quality (RSQ\%) for different quantum optimization methods across Maximum Independent Set (MIS) instances. The instance 1tc.8 is omitted for clarity. PCE consistently achieves high RSQ\%, while QRAO and QAOA exhibit greater variability. The results highlight the trade-off between solution quality and method choice.}

    \label{fig:mis_rsq_compare}
\end{figure}

\subsection{QAP Results}

The performance of \textbf{VQE}, \textbf{PCE}, and \textbf{CVaR VQE} on classically benchmarked Quadratic Assignment Problem (QAP) instances is summarized in Table~\ref{tab:results_qap}. The evaluation focuses on three key aspects: \textit{solution feasibility, optimality gap, and qubit efficiency}.

\paragraph{Feasibility Across Methods}
\begin{itemize}
    \item VQE and CVaR VQE fail to produce feasible solutions for all QAP instances, indicating challenges in convergence and optimization.
    \item PCE successfully finds feasible solutions across all instances.
\end{itemize}

\paragraph{Optimality Gap and Solution Quality}
\begin{itemize}
    \item While PCE finds feasible solutions, it exhibits large optimality gaps, ranging from $15.47\%$ (chr12b) to $234.75\%$ (chr12a), highlighting limitations in solution accuracy.
    \item No results are available for VQE and CVaR VQE due to infeasibility.
\end{itemize}

\paragraph{Qubit Efficiency and Scalability}
\begin{itemize}
    \item PCE significantly reduces qubit requirements to just $11$ qubits per instance, compared to the full QUBO encoding required by VQE.
\end{itemize}

\paragraph{Key Takeaways}
\begin{itemize}
    \item PCE is the only method that produces feasible solutions, making it the best-performing approach among those tested.
    \item VQE and CVaR VQE fail entirely on these instances, indicating that these variational methods may not be well-suited for QAP.
    \item Despite its feasibility, PCE exhibits high optimality gaps. This challenge likely stems from QAP’s dense interaction structure and the rapid growth of possible assignments as the problem size increases, both of which are particularly demanding for current quantum algorithms. Consequently, addressing QAP effectively may require deeper circuits or novel algorithmic strategies to manage its inherent complexity.
    \item Qubit-efficient encodings like PCE are essential for making QAP solvable on near-term quantum devices, even if solution quality remains a challenge.
\end{itemize}

\begin{figure}
    \centering
    \includegraphics[width=\linewidth]{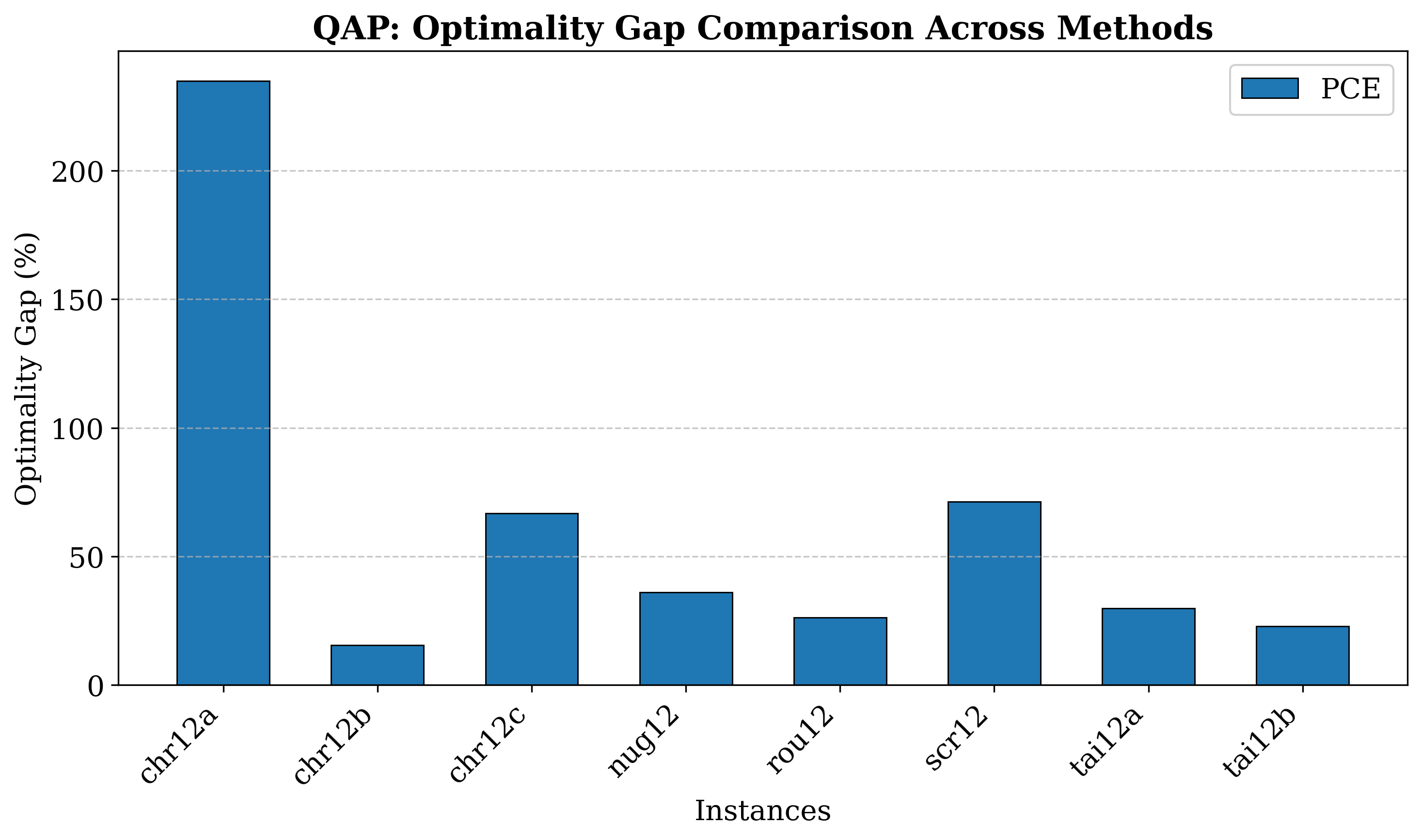}
    \caption{Optimality gap comparison for Quadratic Assignment Problem (QAP) instances using Pauli Correlation Encoding (PCE). The results highlight variations in gap performance, with some instances showing significant deviation from optimal solutions. These insights help evaluate the feasibility of quantum approaches for combinatorial optimization.}

    \label{fig:qap_plot}
\end{figure}

\begin{table*}
    \caption{\label{tab:results_qap}Performance comparison of Variational Quantum Eigensolver (VQE), Pauli Correlation Encoding (PCE), and Quantum Random Access Optimization (QRAO) on classically benchmarked hard instances of the Quadratic Assignment Problem (QAP). The "Variables" column indicates the number of variables in the quadratic unconstrained binary optimization (QUBO) formulation. The columns under "VQE", "PCE", and "QRAO" report the number of qubits used, whether a feasible solution was found (Feasibility: Yes/No), and the optimality gap (\%). Results highlight the feasibility and effectiveness of different quantum approaches in solving QAP instances.}
    
    \begin{ruledtabular}
    \begin{tabular*}{\textwidth}{@{\extracolsep{\fill}}c|c|c|cc|ccc|cc}
        \textbf{Instance} & \textbf{Optimal} & \textbf{Variables} & 
        \multicolumn{2}{c|}{\textbf{VQE}} & 
        \multicolumn{3}{c|}{\textbf{PCE}} & 
        \multicolumn{2}{c}{\textbf{CVaR VQE}} \\
        \cline{4-10}
        & & &  \textbf{Feas} & \textbf{Gap (\%)} & 
        \textbf{Qubits} & \textbf{Feas} & \textbf{Gap (\%)} & 
         \textbf{Feas} & \textbf{Gap (\%)} \\
        \hline
        chr12a & 9552 & 144 &  No & -- & 11 & Yes & 234.75 &  No & -- \\
        chr12b & 9742 & 144 &  No & -- & 11 & Yes & 15.47 &  No & -- \\
        chr12c & 11156 & 144 &  No & -- & 11 & Yes & 66.82 &  No & -- \\
        nug12  & 578 & 144 &  No & -- & 11 & Yes & 35.98 &  No & -- \\
        rou12  & 235528  & 144 &  No & -- & 11 & Yes & 26.30 &  No & -- \\
        scr12 & 31410  & 144 &  No & -- & 11 & Yes & 71.30 & No & -- \\
        tai12a & 224416 & 144 & No & -- & 11 & Yes & 29.78 &  No & -- \\
        tai12b & 39464925 & 144 &  No & -- & 11 & Yes & 22.90 &  No & -- \\
    \end{tabular*}
    \end{ruledtabular}
\end{table*}

\subsection{MSP Results}

The performance of \textbf{PCE}, \textbf{VQE}, and \textbf{CVaR VQE} on Market Share Problem (MSP) instances is summarized in Table~\ref{tab:results_ms}. The evaluation focuses on \textit{solution feasibility, constraint satisfaction, and solution quality}.

\paragraph{Feasibility and Constraint Violations}
\begin{itemize}
    \item PCE produces feasible solutions in three instances (2x10 S0, 2x10 S1, and 5x40 S0), demonstrating its effectiveness in handling MSP constraints.
    \item VQE and CVaR VQE fail to find feasible solutions in all instances, with significant constraint violations.
    \item Constraint violations in PCE are limited to one or two constraints per instance, whereas VQE and CVaR VQE exhibit higher magnitudes of violations across multiple constraints.
\end{itemize}

\paragraph{Solution Quality}
\begin{itemize}
    \item PCE achieves the best solution quality, consistently yielding the lowest objective values, which is desirable in a minimization problem.
    \item VQE and CVaR VQE produce significantly higher objective values, indicating suboptimal performance. Additionally, their high constraint violations further reduce solution validity.
\end{itemize}

\paragraph{Qubit Efficiency and Scalability}
\begin{itemize}
    \item PCE requires significantly fewer qubits (ranging from 7 to 11), making it a more hardware-efficient encoding method.
    \item VQE and CVaR VQE inherently require full QUBO encoding, leading to higher qubit requirements and scalability concerns.
\end{itemize}

\paragraph{Key Takeaways}
\begin{itemize}
    \item PCE is the only method capable of producing feasible solutions, reinforcing its superiority for MSP with constraint satisfaction.
    \item VQE and CVaR VQE struggle with feasibility, making them less reliable for solving MSP in their current formulation.
    \item Constraint adherence is critical in MSP, and PCE demonstrates better trade-offs between feasibility and solution quality.
    \item Qubit-efficient encoding techniques like PCE remain essential for practical quantum optimization on near-term hardware.
\end{itemize}

\begin{figure}
    \centering
    \includegraphics[width=\linewidth]{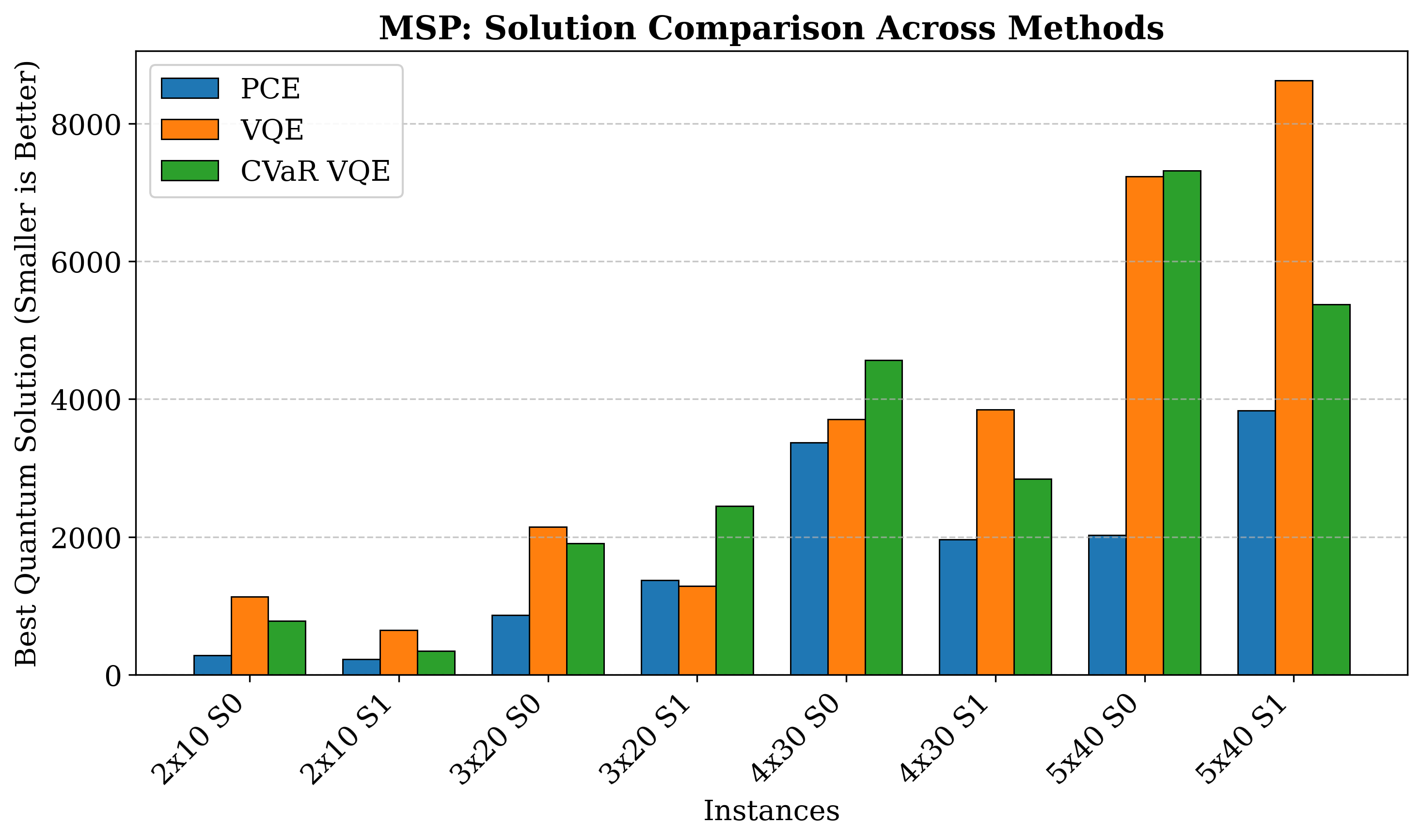}
    \caption{Comparison of best quantum solutions obtained across different methods for Market Share Problem (MSP) instances. Since MSP is a minimization problem, lower values indicate better performance. PCE consistently yields lower solution values, prioritizing feasibility, while VQE and CVaR VQE tend to produce higher values, often with constraint violations. The trend highlights the trade-off between feasibility and solution quality.}

    \label{fig:ms_plot}
\end{figure}

\begin{table*}[t]
    \caption{\label{tab:results_ms}Performance comparison of PCE, VQE, and CVaR VQE on MSP instances. The table reports the number of qubits used (for PCE), feasibility status (Feas: Yes/No), the best quantum solution obtained (Sol.), and constraint violations (Const. Viol.), where the violation magnitude and the number of violated constraints (in brackets) are specified. This comparison highlights the feasibility, solution quality, and constraint adherence of different quantum optimization approaches.}
    \begin{ruledtabular}
    \begin{tabular*}{\textwidth}{@{\extracolsep{\fill}}c|c|c|cccc|ccc|ccc}
        \textbf{Instance} & \textbf{Opt.} & \textbf{Vars.} & 
        \multicolumn{4}{c|}{\textbf{PCE}} & 
        \multicolumn{3}{c|}{\textbf{VQE}} & 
        \multicolumn{3}{c}{\textbf{CVaR VQE}} \\ 
        \cline{4-13}
        & & & \textbf{Qubits} & \textbf{Feas} & \textbf{Sol.} & \textbf{Const. Viol.} & 
        \textbf{Feas} & \textbf{Sol} & \textbf{Const. Viol.} & 
        \textbf{Feas} & \textbf{Sol} & \textbf{Const. Viol.} \\ 
        \hline
        2x10 S0 & 3 & 50 & 7  & Yes & 280 & 0 (0) & No& 1131 & 218 (2) & No & 780 & 372 (2)  \\ 
        2x10 S1 & 3 & 46 & 7  & Yes & 226 & 0 (0) & No & 646 & 199 (2) & No & 341  & 141 (2)  \\
        3x20 S0 & 3 & 84 & 8  & No & 867 & 1 (1) & No & 2144 & 1013 (3) & No  & 1905  & 1154 (3)  \\
        3x20 S1 & 2 & 80 & 8  & No & 1372 & 2 (1) & No & 1283 & 1046 (3) & No  & 2445  & 1368 (3)   \\
        4x30 S0 & 1 & 118 & 10  & No & 3370 & 7 (1) & No  & 3706  & 2150(4)  & No  & 4564  & 2544 (4)   \\
        4x30 S1 & 0 & 118 & 10  & No & 1963 & 6 (2) & No  & 3849  & 920 (4)  & No  & 2841  & 1078 (4)   \\
        5x40 S0 & 1 & 154 & 11  & Yes & 2024 & 0 (0) & No  & 7233  & 3155 (5) & No  & 7322  & 3405 (5)   \\
        5x40 S1 & 1 & 152 & 11  & No & 3833 & 1 (1) & No  & 8626  & 1645 (5)& No  & 5377  & 2034 (5)   \\
    \end{tabular*}
    \end{ruledtabular}
\end{table*}

\section{\label{sec:level9}Conclusion}

Our benchmarking study evaluates a range of quantum optimization techniques on combinatorial optimization problems, including the Multi-Dimensional Knapsack Problem (MDKP), Maximum Independent Set (MIS), Quadratic Assignment Problem (QAP), and Market Share Problem (MSP). By systematically comparing variational methods, qubit-efficient encodings, and classical solvers, we provide insights into the feasibility, optimality gaps, and scalability of quantum approaches.

\subsection{Key Findings}

\paragraph{Pauli Correlation Encoding (PCE) Outperforms Variational Methods in Feasibility}  
PCE consistently produces feasible solutions across MDKP, MIS, QAP, and MSP instances, making it a promising approach. However, part of PCE’s success is attributed to a classical post-processing step, such as bit swap search, which refines solutions beyond what is obtained purely from quantum optimization. In contrast, variational methods like VQE and QAOA often fail to find feasible solutions, particularly in constrained optimization problems.

\paragraph{Qubit Efficiency is a Critical Factor}  
PCE significantly reduces qubit requirements (often using only 7-11 qubits), making it practical for near-term quantum devices. Methods like QRAO and QAOA require significantly more qubits, which limits their scalability on current quantum hardware.

\paragraph{Solution Quality Varies Across Methods}  
In MDKP and MIS, PCE achieves competitive or superior solution quality compared to variational methods. In QAP, while PCE finds feasible solutions, the optimality gap remains high, indicating further refinement is needed. In MSP, PCE outperforms other methods in terms of both feasibility and constraint satisfaction, though obtaining any feasible solution remains challenging.

\paragraph{Variational Methods Struggle with Constraints and Computational Cost}  
While powerful in theory, variational methods like VQE and CVaR VQE exhibit significant constraint violations in MSP and fail to achieve feasible solutions in QAP. Similarly, QAOA and MA-QAOA struggle with feasibility in MIS, limiting their applicability to combinatorial problems with hard constraints. Furthermore, QAOA and its variants were computationally expensive, making them impractical for large problem instances, leading to their exclusion from the largest benchmarks. Notably, even classical solvers like CPLEX faced challenges, failing to provide an optimal solution within the time limit for MSP instances larger than $6 \times 50$ (see Table \ref{tab:cplex_performance_metrics}), highlighting the intrinsic complexity of these problems.

\paragraph{Benchmarking Framework Highlights Scalability Challenges}  
The benchmarking results underscore the importance of problem selection, encoding techniques, and performance metrics. Scalability remains a key challenge, as even on small-to-medium-sized instances, quantum methods often struggled to find feasible or optimal solutions, indicating that large-scale applicability is still far off. Notably, some problem types, such as QAP and MSP, proved particularly difficult, with MSP being the hardest problem for obtaining feasible solutions. These findings further highlight the necessity of systematic benchmarking to guide future improvements in quantum optimization.

\subsection{Future Directions}

\begin{itemize}
    \item \textbf{Noise and Hardware Testing: } While our experiments were conducted on simulators, real quantum hardware introduces additional challenges such as gate errors, decoherence, and connectivity constraints. A crucial next step is to evaluate how these quantum optimization methods perform under real-world conditions. This includes testing algorithms on physical quantum devices to analyze their resilience to noise and determining whether error mitigation techniques, such as zero-noise extrapolation or probabilistic error cancellation, can improve solution quality. Additionally, incorporating realistic noise models into simulations would help bridge the gap between idealized performance and practical feasibility, providing a more accurate assessment of scalability and robustness for near-term quantum devices. 
    \item \textbf{Improved Encoding Strategies:} Further refinements to PCE and other compression techniques could enhance solution quality while maintaining qubit efficiency.
    \item \textbf{Hybrid Quantum-Classical Approaches:} Leveraging classical solvers for pre-processing and post-processing may improve the overall performance of quantum optimization. The use of efficient heuristics can aid in simplifying complex optimization problems, either by reducing the problem size before solving or refining solutions after quantum optimization, thereby improving overall performance.
    \item \textbf{Advanced Parameter Optimization:} Variational methods require better initialization and parameter tuning strategies to improve feasibility and solution quality.
    \item \textbf{Hardware-Specific Optimization:} Optimizing quantum algorithms for specific hardware architectures could improve their practical applicability.
    \item \textbf{Expanding to Other Hard Problems:} Additional hard combinatorial optimization problems, such as the Low Autocorrelation Binary Sequence (LABS) problem and Sports Timetabling problems, can be implemented and tested within the benchmarking framework.
    
\end{itemize}

This study seeds future research in quantum optimization benchmarking, enabling systematic comparisons across problem domains and methodologies. While quantum optimization methods show promise, their success remains highly dependent on problem structure, encoding choices, and the ability to integrate techniques for improving solution quality. Addressing these challenges will be critical for advancing practical quantum optimization in the near term.

\section{Acknowledgment}

This research is supported by the National Research Foundation, Singapore under its Quantum Engineering Programme 2.0 (NRF2021-QEP2-02-P01).

\bibliography{reference}

\begin{thebibliography}{10}

\bibitem{yu2000robust}
Chian-Son Yu and Han-Lin Li.
\newblock A robust optimization model for stochastic logistic problems.
\newblock {\em International journal of production economics}, 64(1-3):385--397, 2000.

\bibitem{zenios1993financial}
Stavros~A Zenios.
\newblock {\em Financial optimization}.
\newblock Cambridge university press, 1993.

\bibitem{resende2008handbook}
Mauricio~GC Resende and Panos~M Pardalos.
\newblock {\em Handbook of optimization in telecommunications}.
\newblock Springer Science \& Business Media, 2008.

\bibitem{SINGH20221636}
Guman Singh and Mohammad Rizwanullah.
\newblock Combinatorial optimization of supply chain networks: A retrospective and literature review.
\newblock {\em Materials Today: Proceedings}, 62:1636--1642, 2022.
\newblock International Conference on Recent Advances in Modelling and Simulations Techniques in Engineering and Science.

\bibitem{karp1972reducibility}
Richard~M. Karp.
\newblock Reducibility among combinatorial problems.
\newblock In {\em Complexity of Computer Computations}, pages 85--103. Plenum Press, 1972.

\bibitem{Abbas_2024}
Amira Abbas, Andris Ambainis, Brandon Augustino, Andreas Bärtschi, Harry Buhrman, Carleton Coffrin, Giorgio Cortiana, Vedran Dunjko, Daniel~J. Egger, Bruce~G. Elmegreen, Nicola Franco, Filippo Fratini, Bryce Fuller, Julien Gacon, Constantin Gonciulea, Sander Gribling, Swati Gupta, Stuart Hadfield, Raoul Heese, Gerhard Kircher, Thomas Kleinert, Thorsten Koch, Georgios Korpas, Steve Lenk, Jakub Marecek, Vanio Markov, Guglielmo Mazzola, Stefano Mensa, Naeimeh Mohseni, Giacomo Nannicini, Corey O’Meara, Elena~Peña Tapia, Sebastian Pokutta, Manuel Proissl, Patrick Rebentrost, Emre Sahin, Benjamin C.~B. Symons, Sabine Tornow, Víctor Valls, Stefan Woerner, Mira~L. Wolf-Bauwens, Jon Yard, Sheir Yarkoni, Dirk Zechiel, Sergiy Zhuk, and Christa Zoufal.
\newblock Challenges and opportunities in quantum optimization.
\newblock {\em Nature Reviews Physics}, 6(12):718–735, October 2024.

\bibitem{classical_benchmark}
Iain Dunning, Swati Gupta, and John Silberholz.
\newblock What works best when? a systematic evaluation of heuristics for max-cut and qubo.
\newblock {\em INFORMS Journal on Computing}, 30:608--624, 08 2018.

\bibitem{farhi2014quantumapproximateoptimizationalgorithm}
Edward Farhi, Jeffrey Goldstone, and Sam Gutmann.
\newblock A quantum approximate optimization algorithm, 2014.

\bibitem{peruzzo2014variational}
Alberto Peruzzo, Jarrod McClean, Peter Shadbolt, Man-Hong Yung, Xiao-Qi Zhou, Peter~J. Love, Al{\'a}n Aspuru-Guzik, and Jeremy~L. O'Brien.
\newblock A variational eigenvalue solver on a photonic quantum processor.
\newblock {\em Nature Communications}, 5:4213, 2014.

\bibitem{Cerezo_2022}
M.~Cerezo, Kunal Sharma, Andrew Arrasmith, and Patrick~J. Coles.
\newblock Variational quantum state eigensolver.
\newblock {\em npj Quantum Information}, 8(1), September 2022.

\bibitem{Zhang_2022}
Yu~Zhang, Lukasz Cincio, Christian F.~A. Negre, Piotr Czarnik, Patrick~J. Coles, Petr~M. Anisimov, Susan~M. Mniszewski, Sergei Tretiak, and Pavel~A. Dub.
\newblock Variational quantum eigensolver with reduced circuit complexity.
\newblock {\em npj Quantum Information}, 8(1), August 2022.

\bibitem{Blekos_2024}
Kostas Blekos, Dean Brand, Andrea Ceschini, Chiao-Hui Chou, Rui-Hao Li, Komal Pandya, and Alessandro Summer.
\newblock A review on quantum approximate optimization algorithm and its variants.
\newblock {\em Physics Reports}, 1068:1–66, June 2024.

\bibitem{chekuri2004multidimensional}
Chandra Chekuri and Sanjeev Khanna.
\newblock On multidimensional packing problems.
\newblock {\em SIAM journal on computing}, 33(4):837--851, 2004.

\bibitem{lawler1980generating}
Eugene~L. Lawler, Jan~Karel Lenstra, and AHG Rinnooy~Kan.
\newblock Generating all maximal independent sets: Np-hardness and polynomial-time algorithms.
\newblock {\em SIAM Journal on Computing}, 9(3):558--565, 1980.

\bibitem{sahni1976p}
Sartaj Sahni and Teofilo Gonzalez.
\newblock P-complete approximation problems.
\newblock {\em Journal of the ACM (JACM)}, 23(3):555--565, 1976.

\bibitem{mdkp}
Hans Kellerer, Ulrich Pferschy, and David Pisinger.
\newblock {\em Multidimensional Knapsack Problems}, pages 235--283.
\newblock Springer Berlin Heidelberg, Berlin, Heidelberg, 2004.

\bibitem{mis}
Michael~R. Garey and David~S. Johnson.
\newblock {\em Computers and Intractability: A Guide to the Theory of NP-Completeness}.
\newblock W. H. Freeman and Company, 1979.

\bibitem{market_share}
Gérard Cornuéjols and Milind Dawande.
\newblock A class of hard small 0-1 programs.
\newblock {\em INFORMS Journal on Computing}, 11(2):205--210, 1999.

\bibitem{qap}
Tjalling~C. Koopmans and Martin Beckmann.
\newblock Assignment problems and the location of economic activities.
\newblock {\em Econometrica}, 25(1):53--76, 1957.

\bibitem{pardalos1994quadratic}
Panos~M. Pardalos, Franz Rendl, and Henry Wolkowicz.
\newblock The quadratic assignment problem: A survey and recent developments.
\newblock In Panos~M. Pardalos and Henry Wolkowicz, editors, {\em Quadratic Assignment and Related Problems}, volume~16 of {\em DIMACS Series in Discrete Mathematics and Theoretical Computer Science}, pages 1--42. American Mathematical Society, Providence, RI, 1994.

\bibitem{packebusch2016low}
Tom Packebusch and Stephan Mertens.
\newblock Low autocorrelation binary sequences.
\newblock {\em Journal of Physics A: Mathematical and Theoretical}, 49(16):165001, 2016.

\bibitem{van2023international}
David Van~Bulck and Dries Goossens.
\newblock The international timetabling competition on sports timetabling (itc2021).
\newblock {\em European Journal of Operational Research}, 308(3):1249--1267, 2023.

\bibitem{sciorilli2025towards}
Marco Sciorilli, Lucas Borges, Taylor~L Patti, Diego Garc{\'\i}a-Mart{\'\i}n, Giancarlo Camilo, Anima Anandkumar, and Leandro Aolita.
\newblock Towards large-scale quantum optimization solvers with few qubits.
\newblock {\em Nature Communications}, 16(1):476, 2025.

\bibitem{Sharma2025QOB}
SMU‑Quantum.
\newblock Quantum optimization benchmarks.
\newblock \url{https://github.com/SMU-Quantum/quantum-optimization-benchmarks}, 2025.
\newblock Accessed: March 19, 2025.

\bibitem{Sharma2025QOA}
SMU-Quantum
\newblock Quantum optimization algorithms.
\newblock \url{https://github.com/SMU-Quantum/quantum-optimization-algorithms}, 2025.
\newblock Accessed: March 19, 2025.

\bibitem{ibm_quantum_roadmap}
{IBM}.
\newblock Ibm quantum roadmap, 2024.
\newblock Accessed: 2025-02-15.

\bibitem{quantinuum_roadmap_2024}
{Quantinuum}.
\newblock Quantinuum unveils accelerated roadmap to achieve universal, fully fault-tolerant quantum computing by 2030, September 2024.
\newblock Accessed: 2025-02-15.

\bibitem{montanez2024unbalanced}
JA~Montanez-Barrera, Dennis Willsch, Alberto Maldonado-Romo, and Kristel Michielsen.
\newblock Unbalanced penalization: A new approach to encode inequality constraints of combinatorial problems for quantum optimization algorithms.
\newblock {\em Quantum Science and Technology}, 9(2):025022, 2024.

\bibitem{takabayashi2024subgradientmethodusingquantum}
Taisei Takabayashi, Takeru Goto, and Masayuki Ohzeki.
\newblock Subgradient method using quantum annealing for inequality-constrained binary optimization problems, 2024.

\bibitem{hadfield2019quantum}
Stuart Hadfield, Zhihui Wang, Bryan O’gorman, Eleanor~G Rieffel, Davide Venturelli, and Rupak Biswas.
\newblock From the quantum approximate optimization algorithm to a quantum alternating operator ansatz.
\newblock {\em Algorithms}, 12(2):34, 2019.

\bibitem{Egger_2021}
Daniel~J. Egger, Jakub Mareček, and Stefan Woerner.
\newblock Warm-starting quantum optimization.
\newblock {\em Quantum}, 5:479, June 2021.

\bibitem{herrman2022multi}
Rebekah Herrman, Phillip~C Lotshaw, James Ostrowski, Travis~S Humble, and George Siopsis.
\newblock Multi-angle quantum approximate optimization algorithm.
\newblock {\em Scientific Reports}, 12(1):6781, 2022.

\bibitem{Nakanishi_2019}
Ken~M. Nakanishi, Kosuke Mitarai, and Keisuke Fujii.
\newblock Subspace-search variational quantum eigensolver for excited states.
\newblock {\em Physical Review Research}, 1(3), October 2019.

\bibitem{Higgott_2019}
Oscar Higgott, Daochen Wang, and Stephen Brierley.
\newblock Variational quantum computation of excited states.
\newblock {\em Quantum}, 3:156, July 2019.

\bibitem{Barkoutsos_2020}
Panagiotis~Kl. Barkoutsos, Giacomo Nannicini, Anton Robert, Ivano Tavernelli, and Stefan Woerner.
\newblock Improving variational quantum optimization using cvar.
\newblock {\em Quantum}, 4:256, April 2020.

\bibitem{fuller2024approximate}
Bryce Fuller, Charles Hadfield, Jennifer~R Glick, Takashi Imamichi, Toshinari Itoko, Richard~J Thompson, Yang Jiao, Marna~M Kagele, Adriana~W Blom-Schieber, Rudy Raymond, et~al.
\newblock Approximate solutions of combinatorial problems via quantum relaxations.
\newblock {\em IEEE Transactions on Quantum Engineering}, 2024.

\bibitem{Sharma_2024}
Monit Sharma, Yan Jin, Hoong~Chuin Lau, and Rudy Raymond.
\newblock Quantum relaxation for solving multiple knapsack problems.
\newblock In {\em 2024 IEEE International Conference on Quantum Computing and Engineering (QCE)}, page 692–698. IEEE, September 2024.

\bibitem{preskill2018quantum}
John Preskill.
\newblock Quantum computing in the nisq era and beyond.
\newblock {\em Quantum}, 2:79, 2018.

\bibitem{kadowaki1998quantum}
Tadashi Kadowaki and Hidetoshi Nishimori.
\newblock Quantum annealing in the transverse ising model.
\newblock {\em Physical Review E}, 58(5):5355, 1998.

\bibitem{farhi2000quantum}
Edward Farhi, Jeffrey Goldstone, Sam Gutmann, and Michael Sipser.
\newblock Quantum computation by adiabatic evolution.
\newblock {\em arXiv preprint quant-ph/0001106}, 2000.

\bibitem{crosson2021prospects}
EJ~Crosson and DA~Lidar.
\newblock Prospects for quantum enhancement with diabatic quantum annealing.
\newblock {\em Nature Reviews Physics}, 3(7):466--489, 2021.

\bibitem{del2013shortcuts}
Adolfo Del~Campo.
\newblock Shortcuts to adiabaticity by counterdiabatic driving.
\newblock {\em Physical review letters}, 111(10):100502, 2013.

\bibitem{kitaev1995quantum}
A~Yu Kitaev.
\newblock Quantum measurements and the abelian stabilizer problem.
\newblock {\em arXiv preprint quant-ph/9511026}, 1995.

\bibitem{brassard2002quantum}
Gilles Brassard, Peter Hoyer, Michele Mosca, and Alain Tapp.
\newblock Quantum amplitude amplification and estimation.
\newblock {\em Contemporary Mathematics}, 305:53--74, 2002.

\bibitem{gacon2020quantum}
Julien Gacon, Christa Zoufal, and Stefan Woerner.
\newblock Quantum-enhanced simulation-based optimization.
\newblock In {\em 2020 IEEE International Conference on Quantum Computing and Engineering (QCE)}, pages 47--55. IEEE, 2020.

\bibitem{sharma2024quantum}
Monit Sharma, Hoong~Chuin Lau, and Rudy Raymond.
\newblock Quantum enhanced simulation based optimization for newsvendor problems.
\newblock In {\em 2024 IEEE International Conference on Quantum Computing and Engineering (QCE)}, page 457–468. IEEE, September 2024.

\bibitem{sharma2024quantummontecarlomethods}
Monit Sharma and Hoong~Chuin Lau.
\newblock Quantum monte carlo methods for newsvendor problem with multiple unreliable suppliers, 2024.

\bibitem{puchinger2010multidimensional}
Jakob Puchinger, G{\"u}nther~R Raidl, and Ulrich Pferschy.
\newblock The multidimensional knapsack problem: Structure and algorithms.
\newblock {\em INFORMS Journal on Computing}, 22(2):250--265, 2010.

\bibitem{dataset}
John Drake.
\newblock Benchmark instances for the multidimensional knapsack problem, 01 2015.

\bibitem{kroger2024polytime}
Samuel Kroger, Hamidreza Validi, and Illya~V Hicks.
\newblock A polytime preprocess algorithm for the maximum independent set problem.
\newblock {\em Optimization Letters}, 18(2):651--661, 2024.

\bibitem{sloane2000challenge}
N.~J.~A. Sloane.
\newblock Challenge problems: Independent sets in graphs.
\newblock \url{https://oeis.org/A265032/a265032.html}, 2000--.
\newblock OEIS Foundation, 11 South Adelaide Avenue, Highland Park, NJ 08904, USA.

\bibitem{qaplib}
R.~E. Burkard, S.~E. Karisch, and F.~Rendl.
\newblock Qaplib -- a quadratic assignment problem library.
\newblock \url{https://coral.ise.lehigh.edu/data-sets/qaplib/}, 1997.
\newblock Accessed: 2025-02-12.

\bibitem{qiskit2024}
Ali Javadi-Abhari, Matthew Treinish, Kevin Krsulich, Christopher~J. Wood, Jake Lishman, Julien Gacon, Simon Martiel, Paul~D. Nation, Lev~S. Bishop, Andrew~W. Cross, Blake~R. Johnson, and Jay~M. Gambetta.
\newblock Quantum computing with {Q}iskit, 2024.

\bibitem{barahona1989experiments}
Francisco Barahona, Michael J{\"u}nger, and Gerhard Reinelt.
\newblock Experiments in quadratic 0--1 programming.
\newblock {\em Mathematical programming}, 44(1):127--137, 1989.

\bibitem{POLJAK198699}
Svatopluk Poljak and Daniel Turzík.
\newblock A polynomial time heuristic for certain subgraph optimization problems with guaranteed worst case bound.
\newblock {\em Discrete Mathematics}, 58(1):99--104, 1986.

\end{thebibliography}
\clearpage
\appendix
\onecolumngrid 

\section{\label{app:performance}Performance Metrics for MSP Instances}

\begin{table*}[h!]
\centering
\caption{Performance Metrics for Various Problem Sizes in the Market Share Problem. 
The table provides key insights into the performance of the CPLEX solver. \textbf{Gap} represents the relative difference between the best-known solution (incumbent) and the best possible solution (lower bound), where values close to $1$ (e.g., $0.999999999$) indicate difficulty in finding a near-optimal solution within the allotted time. \textbf{Num Nodes} denotes the total number of nodes explored in the branch-and-bound tree, each representing a subproblem. \textbf{Iterations} refer to the total number of iterations performed by the simplex or barrier algorithm during optimization. \textbf{Problem Size} indicates the dimensionality of the problem, expressed as $m \times n$ (products $\times$ retailers). \textbf{Solution} is the best feasible solution found within the time limit, and \textbf{Time} represents the total computation time for each instance in seconds. For problem sizes ranging from $6 \times 50$ to $10 \times 90$, a time limit of one hour was imposed per instance.}

\label{tab:cplex_performance_metrics}
\begin{tabular}{@{}cccccc@{}}
\toprule
\textbf{Problem Size} & \textbf{Num Nodes} & \textbf{Iterations} & \textbf{Gap} & \textbf{Solution} & \textbf{Time(sec.)} \\ \midrule
$3 \times 20$ & 7,552  & 9,776  & 0.0  & 3 & 2.97 \\
$3 \times 20$ & 7,270  & 9,035 & 0.0 & 2 & 1.70 \\
$3 \times 20$ & 7,559  & 10,287 & 0.0 & 3 & 2.05 \\
$3 \times 20$ & 9,205  & 11,446 & 0.0 & 3 & 1.13 \\
$3 \times 20$ & 8,301  & 11,358 & 0.0 & 2 & 1.47 \\ \midrule
$4 \times 30$ & 765,655  & 1,622,772 & 0.0  & 1 & 5.64 \\
$4 \times 30$ & 72,436  & 138,158 & 0.0 & 0 & 5.92\\
$4 \times 30$ & 439,294  &  960,557 & 0.0 & 2 & 7.33\\
$4 \times 30$ & 538,857  & 1,197,837 & 0.0 & 1 & 9.28\\
$4 \times 30$ & 772,561  & 1,687,113 & 0.0 & 1 & 7.33\\ \midrule
$5 \times 40$ & 50,288,061  & 117,363,318  & 0.0  & 1  & 777.33 \\
$5 \times 40$ & 72,691,565  & 173,591,038 & 0.0 & 1 &1076.48 \\
$5 \times 40$ & 65,010,492  & 153,229,808 & 0.0 & 1 &776.88 \\
$5 \times 40$ & 41,687,999  & 98,461,473 & 0.0 & 1 & 291.98 \\
$5 \times 40$ & 14,683,912  & 36,843,546 & 0.0 & 0 & 94.52\\ \midrule
$6 \times 50$ & 263,122,381  & 68,237,5105  & 0.999999999  & 1 & 3607.04 \\
$6 \times 50$ & 433,287,296  & 1,046,486,269 & 0.9999999995 & 2 &3600.30 \\
$6 \times 50$ & 426,949,570  & 1,037,766,414 & 0.9999999995 & 2 & 3600.33 \\
$6 \times 50$ & 544,066,780  & 1,327,364,155 & 0.9999999995 & 2 & 3600.37\\
$6 \times 50$ & 477,967,274  & 1,147,025,106 & 0.9999999995 & 2 & 3600.24\\ \midrule
$7 \times 60$ & 448,161,278  & 1,166,369,595 & 0.9999999998 & 4 & 3616.47 \\
$7 \times 60$ & 344,532,350  & 910,970,829  & 0.9999999998 & 6 & 3616.81\\
$7 \times 60$ & 409,426,464  & 1,086,559,026 & 0.9999999998 & 5 &  3617.24\\
$7 \times 60$ & 491,680,008  & 1,327,295,105 & 0.9999999998 & 5 & 3619.41\\
$7 \times 60$ & 426,054,792  & 1,120,381,222 & 0.9999999998 & 6 & 3616.79\\ \midrule
$8 \times 70$ & 295,800,675  & 855,015,623  & 0.9999999999 & 10 & 3614.97\\
$8 \times 70$ & 353,441,910  & 1,031,585,153 & 0.9999999999 & 9 & 3615.11\\
$8 \times 70$ & 475,771,200  & 1,409,513,865 & 0.9999999999 & 10 & 3603.86\\
$8 \times 70$ & 347,236,031  & 1,000,525,057 & 0.9999999998 & 4 & 3616.90\\
$8 \times 70$ & 443,654,750  & 1,265,304,303 & 0.9999999999 & 8 & 3617.91\\ \midrule
$9 \times 80$ & 334,649,890  & 1,057,697,331 & 0.9999999999 & 13 & 3605.13\\
$9 \times 80$ & 96,145,990   & 313,334,985  & 0.9999999999 & 16 & 3625.61\\
$9 \times 80$ & 34,232,455   & 108,912,889  & 0.9999999999 & 21 & 3665.30\\
$9 \times 80$ & 370,702,271  & 1,192,690,111 & 0.9999999999 & 12 & 3605.00\\
$9 \times 80$ & 350,650,734  & 1,117,869,759 & 0.9999999999 & 15 & 3604.72\\ \midrule
$10 \times 90$ & 44,257,054   & 151,067,766  & 0.999999999  & 31 & 3714.82\\
$10 \times 90$ & 22,423,7339  & 791,966,872  & 0.9999999900 & 20 & 3611.07\\
$10 \times 90$ & 15,784,9669  & 571,310,599  & 0.9999999999 & 23 & 3620.97\\
$10 \times 90$ & 42,569,893   & 151,946,948  & 0.9999999999 & 26 & 3710.53\\
$10 \times 90$ & 170,802,613  & 603,450,591  & 0.9999999999 & 22 & 3614.84\\ \bottomrule
\end{tabular}

\end{table*}
\clearpage

\twocolumngrid

\section{\label{app:pce} Pauli Correlation Optimization} 

Current quantum hardware imposes significant limitations on the size of QUBO problems that can be addressed. The treatable problem size is constrained by the hardware's physical capacity and the problem's density and complexity. As a result, many problems must be simplified or scaled down to fit within these limitations, often at the expense of solution accuracy or completeness.

To fully leverage quantum computing's potential, it is crucial to develop advanced algorithmic techniques for reducing problem size without sacrificing the quality of the solutions. This may involve pre-processing steps, decomposition methods, or hybrid classical-quantum approaches that strategically break down large problems into smaller, more manageable sub-problems. By addressing these challenges, the field can push the boundaries of what is achievable with quantum computing, paving the way for practical applications in solving large-scale combinatorial optimization problems.

In this work, we introduce two key advancements to the Pauli Correlation Encoding (PCE) method. The first involves replacing the Weighted Max-Cut loss function with a QUBO-based formulation, and the second incorporates a multiple re-optimization strategy combined with multi-bit swap. These enhancements can be implemented independently or in conjunction with each other

The first enhancement broadens the applicability of PCE, making it a more general-purpose method. This can also be achieved by transforming any QUBO formulation into an equivalent Weighted Max-Cut problem, since it is well-established that any QUBO problem can be translated into an equivalent weighted Max-Cut problem; more precisely, a QUBO problem defined on $n$ variables can be transformed into a Max-Cut problem on $n+1$ vertices \cite{barahona1989experiments}.  we provide code implementations for both approaches. The second enhancement introduces perturbations to the trained parameters whenever the optimization process encounters local minima, thereby reducing the risk of premature convergence.

Upon reaching a local minimum, an exhaustive local search is performed using single or multi-bit swap. The resulting parameters are then perturbed, followed by a new minimization phase until convergence is achieved. If the subsequent local search results in an improved cut value, the parameters are updated, and the process is repeated. If no improvement is observed after $N$ perturbations, the optimization procedure is terminated.

PCE represents a novel methodology for tackling combinatorial optimization problems involving $m = \mathcal{O}(n^k)$ binary variables using only $n$ qubits, where $k$ is a selected integer. This technique achieves dimensionality reduction by mapping the binary variables onto $m$ Pauli matrix correlations distributed across multiple qubits.

The fundamental principle of this approach involves encoding binary variables in the optimization problem into $k$-body Pauli operators. In contrast, traditional encodings within the Quantum Approximate Optimization Algorithm (QAOA) utilize a single-qubit $Z$ operator for each binary variable.

In this method, the $m$ binary variables are encoded into $m$ Pauli correlations following $m = \mathcal{O}(n^k)$, where $k$ is an integer determined by our choice, and $n$ is the number of qubits.

We encode the binary variables $x = \{x_i\}_{i \in [m]}$ in relation to a specific subset of Pauli strings $\Pi_i$, excluding the $n$-fold tensor product of the identity operator, using:

\begin{equation} 
    x_i = \text{sgn} (\langle \Pi_i \rangle ) \text{ for all } i \in [m] 
    \label{eq:sgn}
\end{equation}

Here, $\text{sgn}$ denotes the sign function, and $\langle \Pi_i \rangle = \langle \Psi | \Pi_i | \Psi \rangle$ represents the expectation value of $\Pi_i$ with respect to the quantum state $|\Psi\rangle$ of the parameterized quantum circuit.

The state in Eq. (\ref{eq:sgn}) is parameterized as the output of a quantum circuit characterized by parameters $\theta$, such that $|\Psi\rangle = |\Psi (\theta)\rangle$. These parameters $\theta$ are optimized using a variational approach. This optimization aims to minimize a non-linear loss function defined as:

\begin{equation}\label{eq
} \mathcal{L} = \sum_{(i,j)\in E} W_{ij}\tanh\big(\alpha \langle \Pi_i\rangle\big)\tanh\big(\alpha \langle \Pi_j\rangle\big) + \mathcal{L}^{(\text{reg})}. \end{equation}

The first term of this loss function corresponds to a relaxation of the binary Max-Cut problem, where the sign functions from Eq. (\ref{eq:sgn}) are replaced by smooth hyperbolic tangent functions, which are more amenable to gradient-based optimization techniques. The second term, $\mathcal{L}^{(\text{reg})}$, serves as a regularization component that drives all correlators towards zero, an approach that has been observed to enhance the performance of the optimization solver.

The regularization term in the loss function penalizes large correlator values, thereby constraining the optimizer to remain within the correlator domain where all possible bit string solutions are accessible. This approach ensures that the optimization process explores a solution space that can fully express the desired bit string configurations.

\begin{equation} \mathcal{L}^{(reg)} = \beta \nu \Bigg[ \frac{1}{m} \sum_{i \in V} \tanh (\alpha \langle \Pi_i \rangle )^2 \Bigg]^2, \end{equation}

where the factor $\frac{1}{m}$ serves to normalize the sum within the square brackets. The parameter $\nu$ represents a lower bound on the weighted Max-Cut value, specifically utilizing the Poljak-Turzík lower bound \cite{POLJAK198699}, given by:

\begin{equation*} \nu = \frac{w(G)}{2} + \frac{w(T_{min})}{4}, \end{equation*}

where $w(G)$ denotes the total weight of the graph, and $w(T_{min})$ is the weight of its minimum spanning tree. The hyperparameter $\beta$ is set as a free parameter, typically fixed at $\beta = \frac{1}{2}$. This regularization strategy not only normalizes the correlator contributions but also leverages structural properties of the graph to guide the optimization, using $\nu$ to incorporate fundamental bounds on the problem. The careful calibration of $\beta$ and $\nu$ thus plays a crucial role in stabilizing the optimization, ensuring that the algorithm remains within a feasible and expressive domain throughout the training process.

After completing the training phase, the circuit’s output state is measured, yielding a bit-string \(x\) as determined by Eq.~(\ref{eq:sgn}). To refine the solution, a multi-phase re-optimization process is initiated by introducing small perturbations to the trained parameters. Multi-phase re-optimization involves iteratively refining the solution through several stages. It begins with an initial optimization phase using Pauli Correlation Encoding (PCE) to obtain a baseline solution. Next, perturbations are introduced to explore different regions of the solution space and avoid local minima. This is followed by an exhaustive local search around the perturbed solution to identify potential improvements.

\section*{QUBO Loss}

For the first enhancement, we utilize the QUBO formulation, which is given by:

\begin{equation}
    \min_{\mathbf{x} \in \{0, 1\}^n} \mathbf{x}^\top Q \mathbf{x} + \mathbf{c}^\top \mathbf{x} + \text{offset},
\end{equation}
where:
\begin{itemize}
    \item \( Q \in \mathbb{R}^{n \times n} \) is a symmetric quadratic cost matrix,
    \item \( \mathbf{c} \in \mathbb{R}^n \) represents linear coefficients,
    \item \( \text{offset} \) is a constant term.
\end{itemize}

and the updated loss function is formulated as follows:

\begin{align}
\mathcal{L} &= \sum_{(i, j) \in E} Q_{ij}\,\tanh\big(\alpha \langle\pi_i\rangle\big)\,\tanh\big(\alpha \langle\pi_j\rangle\big) \notag\\[1mm]
&\quad + \sum_{i=1}^m c_i\,\Bigl(\tanh\big(\alpha \langle\pi_i\rangle\big)\Bigr)^2 + \mathcal{L}^{(reg)}
\end{align}

where the regularization loss is 

\begin{equation}
        \mathcal{L}^{(reg)}= \beta \nu \cdot \left[ \frac{1}{m} \sum_{i=1}^m \left( \tanh(\alpha \langle\pi_i\rangle) \right)^2 \right]^2.
\end{equation}

Since we are not dealing with Max-Cut or Weighted Max-Cut, the parameter $\nu$ does not require a lower bound. Instead, we use the Frobenius norm of $Q$
    \begin{equation}
        \nu = c \cdot \sqrt{\sum_{i,j} Q_{ij}^2},
    \end{equation}

The quantum ansatz \( \psi(\boldsymbol{\theta}) \) is constructed using a parameterized \texttt{Brickwork} circuit, which consists of the following components:
\begin{itemize}
    \item Alternating layers of single-qubit rotation gates \( R_X, R_Y, R_Z \),
    \item Entangling $R_{XX}$ arranged in a brickwork pattern to ensure connectivity among qubits,
\end{itemize}

This quantum circuit represents the Brickwork ansatz for a quantum system with \(n\) qubits and depth \(d\). The ansatz alternates between parameterized single-qubit rotation gates and entangling layers.

\begin{figure}[h]
    \centering
    \resizebox{0.5\textwidth}{!}{
        \begin{quantikz}
            \lstick{\ket{0}}&\gate{R_X(\theta_1)}& \gate[2]{R_{XX}(\theta_5)}&\gate{R_Y(\theta_7)}&& \gate{R_Z(\theta_{12})}&\gate[2]{R_{XX}(\theta_{16})}& \\
            \lstick{\ket{0}}&\gate{R_X(\theta_2)}&&\gate{R_Y(\theta_8)}&\gate[2]{R_{XX}(\theta_{11})}&\gate{R_Z(\theta_{13})}&& \\
            \lstick{\ket{0}}&\gate{R_X(\theta_3)}&\gate[2]{R_{XX}(\theta_6)}&\gate{R_Y(\theta_9)}&& \gate{R_Z(\theta_{14})}& \gate[2]{R_{XX}(\theta_{17})}& \\
            \lstick{\ket{0}}&\gate{R_X(\theta_4)}&&\gate{R_Y(\theta_{10})}&&\gate{R_Z(\theta_{15})}&& 
        \end{quantikz}
    }
    \sloppy
    \caption{Variational ansatz as a brickwork architecture, with layers of single-qubit rotations around a single direction ($X$, $Y$, or $Z$), one at a time, and a layer of two-qubit rotation gates ($R_{XX}$). The circuit depicts three complete layers of the BrickWork ansatz.}
    \fussy
    \label{fig:quantum_circuit}
\end{figure}
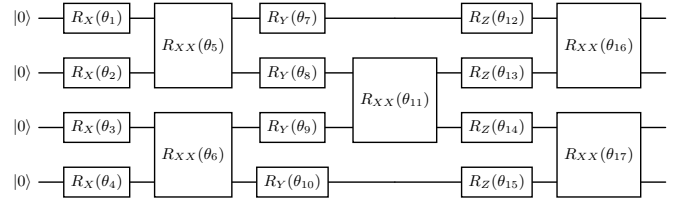

\section*{Dynamic Perturbation and Multi-Reoptimization}

In our optimization framework, we iteratively refine a set of parameters \(\theta\) for solving a QUBO problem. The approach is designed to balance local refinement (exploitation) and global exploration, thereby increasing the likelihood of escaping local minima. The process can be described in the following steps:

\subsection*{1. Initialization}
We begin with an initial parameter set \(\theta\), an initial perturbation factor \(P\), and set the best known parameters \(\theta^*\) with the corresponding QUBO cost \(Q^* = \infty\). A failure counter \(f\) is initialized as \(f = 0\) to track consecutive iterations without improvement. Additionally, a historical trend \(T\) (initialized to a zero vector) is maintained to record the direction of previous parameter updates.

\subsection*{2. Dynamic Perturbation Scaling}
At iteration (or round) \(r\), the perturbation factor is adjusted to facilitate exploration:
\begin{itemize}
    \item Every third round, the perturbation factor is amplified:
    \[
    P' = E \cdot P,
    \]
    where \(E\) is the exploration factor.
    \item Otherwise, \(P' = P\).
\end{itemize}

\subsection*{3. Adaptive Perturbation Application}
The current parameters \(\theta\) are perturbed to generate a candidate \(\tilde{\theta}\) as follows:
\[
\tilde{\theta} = \theta + \Delta_r + \Delta_d,
\]
where
\begin{itemize}
    \item \textbf{Random Perturbation:}
    \[
    \Delta_r \sim \mathcal{N}\Bigl(0,\; \bigl[P'(1 + f/5)\bigr]^2 I\Bigr),
    \]
    which scales the variance of the noise with both the adjusted perturbation factor \(P'\) and the failure count \(f\).
    \item \textbf{Directional Perturbation:}
    \[
    \Delta_d = P' \cdot \operatorname{sgn}(T),
    \]
    which biases the search in the direction suggested by the historical trend \(T\).
\end{itemize}

\subsection*{4. Local Optimization and QUBO Evaluation}
Starting from the perturbed parameters \(\tilde{\theta}\), a local optimization method (e.g., Nelder--Mead) is employed to obtain an optimized set:
\[
\theta_{\text{opt}} = \arg\min_{\theta} Q(\theta),
\]
where \(Q(\theta)\) denotes the QUBO cost function. The optimized parameters are then embedded into the quantum circuit ansatz, and the resultant state is evaluated to compute the cost \(Q(\theta_{\text{opt}})\).

\subsection*{5. Update and Adaptation}
The algorithm then compares \(Q(\theta_{\text{opt}})\) with the best cost \(Q^*\) found so far. If an improvement is detected:
\[
\text{if } Q(\theta_{\text{opt}}) < Q^*:
\]
\[
\begin{aligned}
\theta^* &\leftarrow \theta_{\text{opt}},\\[1mm]
Q^* &\leftarrow Q(\theta_{\text{opt}}),\\[1mm]
f &\leftarrow 0,\\[1mm]
P &\leftarrow P \cdot \delta,
\end{aligned}
\]
where \(\delta < 1\) is a decay factor used to gradually reduce the perturbation magnitude as the search converges.

If no improvement is detected, the failure counter is incremented, \(f \leftarrow f + 1\). The parameters \(\theta\) are then updated based on the level of stagnation:
\begin{itemize}
    \item \textbf{Random Restart:} If \(f\) exceeds a restart threshold \(f_{\text{restart}}\), the parameters are reinitialized uniformly:
    \[
    \theta \sim U(-\pi, \pi).
    \]
    \item \textbf{Weighted Blending:} If \(f\) is even (and below the restart threshold), the new parameters are computed as a weighted blend between the best parameters and a direction informed by \(T\):
    \[
    \theta = \text{weighted\_blend}(\theta^*, T).
    \]
    \item \textbf{Stronger Local Perturbation:} Otherwise, a stronger perturbation is applied:
    \[
    \theta = \text{adaptive\_perturbation}\Bigl(\theta^*,\, 2P,\, f,\, T\Bigr).
    \]
\end{itemize}

\subsection*{6. Checkpointing and Termination}
To ensure progress is not lost, checkpoints containing \(\theta^*\), \(Q^*\), and the current round number are saved periodically. The iterative process continues until the number of consecutive rounds without improvement reaches a predetermined limit.

\bigskip

This framework allows the optimization process to adaptively modulate the balance between exploration and exploitation. By scaling the perturbation factor in response to the failure count and historical trends, the method is better equipped to escape local minima and converge towards a more optimal solution for the QUBO problem.

\onecolumngrid

\begin{algorithm}[H]
\caption{Multi-Step PCE Optimization with Dynamic Perturbations}
\begin{algorithmic}[1]
\Require QUBO problem \( (Q, \mathbf{c}, \text{offset}) \), initial parameters \(\boldsymbol{\theta}_0\), and hyperparameters:
\begin{itemize}
    \item \(P\): initial perturbation factor,
    \item \(E\): exploration factor,
    \item \(\delta\): decay factor,
    \item \(f_{\text{restart}}\): restart threshold,
    \item \(\text{max\_no\_improvement\_rounds}\): maximum allowed rounds without improvement.
\end{itemize}
\Ensure Optimized parameters \(\boldsymbol{\theta}^*\) and best QUBO cost \(Q^*\)
\State Initialize \(\boldsymbol{\theta} \gets \boldsymbol{\theta}_0\), \(Q^* \gets \infty\), \(f \gets 0\) \Comment{\(f\) is the failure count}, and set historical trend \(T \gets \mathbf{0}\)
\State Set round counter \(r \gets 0\)
\While{\(f < \text{max\_no\_improvement\_rounds}\)}
    \State \(r \gets r + 1\)
    \If{\(r \mod 3 = 0\)}
        \State Set \(P' \gets E \cdot P\) \Comment{Amplify perturbation every third round}
    \Else
        \State Set \(P' \gets P\)
    \EndIf
    \State \textbf{Adaptive Perturbation:}
    \[
    \tilde{\boldsymbol{\theta}} \gets \boldsymbol{\theta} + \Delta_r + \Delta_d,
    \]
    where
    \[
    \Delta_r \sim \mathcal{N}\Bigl(0,\; \Bigl[P'\Bigl(1 + \frac{f}{5}\Bigr)\Bigr]^2 I\Bigr)
    \quad \text{and} \quad
    \Delta_d = P' \cdot \operatorname{sgn}(T).
    \]
    \State \textbf{Local Optimization:} Compute
    \[
    \boldsymbol{\theta}_{\text{opt}} \gets \arg\min_{\boldsymbol{\theta}} Q(\boldsymbol{\theta}) \quad \text{starting from } \tilde{\boldsymbol{\theta}}.
    \]
    \State Evaluate the cost \(q \gets Q(\boldsymbol{\theta}_{\text{opt}})\).
    \If{\(q < Q^*\)}
        \State Update best solution: \(\boldsymbol{\theta}^* \gets \boldsymbol{\theta}_{\text{opt}}\) and \(Q^* \gets q\).
        \State Reset failure count: \(f \gets 0\).
        \State Decay perturbation: \(P \gets P \cdot \delta\).
    \Else
        \State Increment failure count: \(f \gets f + 1\).
        \If{\(f \ge f_{\text{restart}}\)}
            \State \textbf{Random Restart:} Reinitialize parameters
            \[
            \boldsymbol{\theta} \sim U(-\pi, \pi).
            \]
        \ElsIf{\(f \mod 2 = 0\)}
            \State \textbf{Weighted Blending:} Set
            \[
            \boldsymbol{\theta} \gets \text{weighted\_blend}(\boldsymbol{\theta}^*, T).
            \]
        \Else
            \State \textbf{Stronger Local Perturbation:} Set
            \[
            \boldsymbol{\theta} \gets \text{adaptive\_perturbation}(\boldsymbol{\theta}^*, 2P, f, T).
            \]
        \EndIf
    \EndIf
    \State \textbf{Update Historical Trend:}
    \[
    T \gets \operatorname{sgn}(\boldsymbol{\theta}_{\text{opt}} - \boldsymbol{\theta}) \cdot \bigl|Q^* - q\bigr|.
    \]
    \If{Improvement remains below a threshold for 5 consecutive rounds}
        \State \textbf{Break}
    \EndIf
    \State Set \(\boldsymbol{\theta} \gets \boldsymbol{\theta}_{\text{opt}}\) \Comment{Next starting point}
\EndWhile
\State \textbf{Return:} \(\boldsymbol{\theta}^*\) and \(Q^*\)
\end{algorithmic}
\end{algorithm}

\section{\label{app:metrics}Quantum Algorithm Metrics}

\begin{table*}[h]
    \caption{Resource usage per instance across different approaches. Each instance's resource consumption is detailed for VQE, CVaR VQE, and PCE. For PCE, the reported time represents the total computation time, including both optimization and post-processing. In case of PCE Total Time (Optimization time + Post Processing Time), since it uses bit swap search as a post processing strategy.}
    \renewcommand{\arraystretch}{1.3}
    \centering
    \footnotesize
    \begin{tabular*}{\textwidth}{@{\extracolsep{\fill}} lccccccc}
        \toprule
        \textbf{Instance} & \textbf{Approach} & \textbf{Qubits} & \textbf{Depth} & \textbf{Gate Count} & \textbf{2-Qubit Gates} & \textbf{Parameters} & \textbf{Execution Time (min)} \\
        \midrule
        \textbf{hp1} 
            & VQE     & 60 & 75 & 836 & 236 & 600 & 351.87 \\
            & CVaR VQE & 60 & 71 & 557  & 177  & 480  & 246.78 \\
            & PCE     & 7 & 336 & 1568  & 336  & 364  & 72.46 (70.06 + 2.4) \\
        \midrule
        \textbf{hp2} 
            & VQE     & 67 & 82 & 934 & 264 & 670 & 395.03 \\
            & CVaR VQE & 67 & 78 & 734  & 198  & 538  & 384.31 \\
            & PCE     & 8 & 384 & 2080  & 448  & 480  & 125.75(121.61 + 4.14) \\
        \midrule
        \textbf{pb1} 
            & VQE     & 59 & 74 & 822 & 232 & 590 & 344.09 \\
            & CVaR VQE & 59 & 70 & 646  & 174  & 472  & 241.58 \\
            & PCE     & 7 & 336 & 1568  & 336  & 364  & 71.03 (68.55 + 2.48) \\
        \midrule
        \textbf{pb2} 
            & VQE     & 66 & 81 & 920 & 260 & 660 & 385.17 \\
            & CVaR VQE & 66 & 77 & 723  & 195  & 528  & 312.18 \\
            & PCE     & 8 & 384 & 2080  & 448  & 480  & 118.98 (115.13 + 3.85) \\
        \midrule
        \textbf{pb4} 
            & VQE     & 45 & 60 & 626 & 176 & 450 & 261.93 \\
            & CVaR VQE & 45 & 56 & 492  & 132  & 360  & 174.36 \\
            & PCE     & 6 & 288 & 1128  & 240  & 264  & 28.08 (27.28 + 0.8) \\
        \midrule
        \textbf{pb5} 
            & VQE     & 116 & 131 & 1620 & 460 & 1160 & 678.59 \\
            & CVaR VQE & 116 & 127 & 1273  & 345  & 928  & 593.73 \\
            & PCE     & 10 & 480 & 3320  & 720  & 760  & 336.52 (307.16 + 29.36) \\
        \midrule
        \textbf{pet2} 
            & VQE     & 99 & 114 & 1382 & 392 & 990 & 578.82 \\
            & CVaR VQE & 99 & 110 & 1086  & 294  & 792  & 467.23 \\
            & PCE     & 9 & 432 & 2664  & 576  & 612  & 151.25 (139.35 + 11.9) \\
        \midrule
        \textbf{pet3} 
            & VQE     & 102 & 117 & 1424 & 404 & 1020 & 596.43 \\
            & CVaR VQE & 102 & 113 & 1119  & 303  & 816  & 565.91 \\
            & PCE     & 9 & 432 & 2664  & 576  & 612  & 187.10 (170.75 + 16.35) \\
        \midrule
        \textbf{pet4} 
            & VQE     & 107 & 122 & 1494 & 424 & 1070 & 625.77 \\
            & CVaR VQE & 107 & 118 & 1174  & 318  & 856  & 517.32 \\
            & PCE     & 9 & 432 & 2664  & 576  & 612  & 175.31 (155.13 + 20.18) \\
        \midrule
        \textbf{pet5} 
            & VQE     & 122 & 137 & 1704 & 484 & 1220 & 713.80 \\
            & CVaR VQE & 122 & 133 & 1339  & 363  & 976  & 576.80 \\
            & PCE    & 10 & 480 & 3320  & 720  & 760  & 345.48 (312.18 + 33.3) \\
        \midrule
        \textbf{pet6} 
            & VQE     & 86 & 101 & 1210 & 340 & 860 & 502.53 \\
            & CVaR VQE & 86 & 97 & 943  & 255  & 688  & 477.53 \\
            & PCE     & 9 & 432 & 2664  & 576  & 612  & 198.3 (180.1 + 18.2) \\
        \midrule
        \textbf{pet7} 
            & VQE     & 100 & 115 & 1396 & 396 & 1000 & 584.69 \\
            & CVaR VQE & 100 & 111 & 1097  & 297  & 800  & 550.14 \\
            & PCE     & 9 & 432 & 2664  & 576  & 612  & 198.61 (177.31 + 21.3) \\
        \bottomrule
    \end{tabular*}
    \label{tab:resource_usage_mdkp}
\end{table*}

\begin{table*}[p]
    \caption{Resource Usage by Instance and Approach. Each instance's resources are detailed for VQE, CVaR VQE, PCE, QRAO, QAOA, MA-QAOA, and CVaR QAOA approaches. In case of PCE Total Time (Optimization time + Post Processing Time), since it uses bit swap search as a post processing strategy.}
    \renewcommand{\arraystretch}{1.3}
    \centering
    \footnotesize
    \begin{tabular*}{\textwidth}{@{\extracolsep{\fill}} lccccccc}
        \toprule
        \textbf{Instance} & \textbf{Approach} & \textbf{Qubits} & \textbf{Depth} & \textbf{Gate Count} & \textbf{2-Qubit Gates} & \textbf{Parameters} & \textbf{Execution Time (s)} \\
        \midrule
        \textbf{1dc.64} 
            & VQE         & 50 & 61 & 547 & 147 & 400 & 231.41 \\
            & CVaR VQE    & 50 & 57 & 398  & 98  & 300  & 184.13     \\
            & PCE         & 7 & 336 & 1568  & 336  & 364  & 31.85 (31.15 + 0.7)     \\
            & QRAO        & 18 & 25 & 142 & 34  & 108 & 132.98 \\
            & QAOA        & 50 & 252 & 1377 & 818 & 559 & M.E   \\
            & MA-QAOA     & 50 & 86 & 559  & 409  & 509  & 3135.20     \\
        \midrule
        \textbf{1tc.16} 
            & VQE         & 16 & 27 & 173 & 45  & 128 & 17.38  \\
            & CVaR VQE    & 16 & 23 & 126  & 30  & 96  & 14.25     \\
            & PCE         & 4 & 192 & 464  & 96  & 112  & 1.86 (1.83 + 0.03)     \\
            & QRAO        & 6  & 13 & 54  & 10  & 34  & 4.33   \\
            & QAOA        & 16 & 30 & 114 & 44  & 70  & 52.35  \\
            & MA-QAOA     & 16 & 12 & 70  & 22  & 54  & 4.26     \\
        \midrule
        \textbf{1tc.32} 
            & VQE         & 32 & 43 & 349 & 93  & 256 & 68.20  \\
            & CVaR VQE    & 32 & 39 & 254  & 62  & 192  & 54.44     \\
            & PCE         & 6 & 288 & 1128  & 240  & 264  & 9.86 (9.78 + 0.08)     \\
            & QRAO        & 13 & 20 & 102 & 24  & 78  & 111.16 \\
            & QAOA        & 32 & 54 & 300 & 136 & 164 & 158.67 \\
            & MA-QAOA     & 32 & 20 & 164  & 68  & 132  & 11.78     \\
        \midrule
        \textbf{1et.64} 
            & VQE         & 62 & 73 & 679 & 183 & 496 & 218.50 \\
            & CVaR VQE    & 62 & 69 & 494  & 122  & 372  & 156.15     \\
            & PCE         & 7 & 336 & 1568  & 336  & 364  & 26.80 (25.05 + 1.75)     \\
            & QRAO        & 24 & 31 & 190 & 46  & 144 & 516.55 \\
            & QAOA        & 62 & 105 & 978 & 528 & 450 & 812.75 \\
            & MA-QAOA     & 62 & 37 & 414  & 210  & 352  & 2668.71     \\
        \midrule
        \textbf{1tc.64} 
            & VQE         & 64 & 75 & 701 & 189 & 512 & 248.15 \\
            & CVaR VQE    & 64 & 71 & 510  & 126  & 384  & 186.63     \\
            & PCE         & 8 & 384 & 2080  & 448  & 480  & 34.70 (33.2 + 1.5)     \\
            & QRAO        & 23 & 30 & 182 & 44  & 138 & 310.11 \\
            & QAOA        & 64 & 105 & 768 & 384 & 384 & 702.44 \\
            & MA-QAOA     & 64 & 32 & 350  & 145  & 286  & 47.85     \\
        \midrule
        \textbf{1tc.8} 
            & VQE         & 8  & 19 & 85  & 21  & 64  & 3.06   \\
            & CVaR VQE    & 8 & 15 & 62  & 14  & 48  & 1.36     \\
            & PCE         & 3 & 144 & 240  & 48  & 60  & 0.632 (0.63+0.002)     \\
            & QRAO        & 4  & 11 & 30  & 6   & 24  & 1.41   \\
            & QAOA        & 8  & 12 & 42  & 12  & 30  & 15.23  \\
            & MA-QAOA     & 8 & 6 & 30  & 6  & 22  & 1.51     \\
        \bottomrule
    \end{tabular*}
    \label{tab:resource_usage_mis}
\end{table*}

\begin{table*}[p]
    \caption{Resource Usage by Instance and Approach. Each instance's resources are detailed for VQE, CVaR VQE, and PCE approaches. In case of PCE Total Time (Optimization time + Post Processing Time), since it uses bit swap search as a post processing strategy.}
    \renewcommand{\arraystretch}{1.3}
    \centering
    \footnotesize
    \begin{tabular*}{\textwidth}{@{\extracolsep{\fill}} lccccccc}
        \toprule
        \textbf{Instance} & \textbf{Approach} & \textbf{Qubits} & \textbf{Depth} & \textbf{Gate Count} & \textbf{2-Qubit Gates} & \textbf{Parameters} & \textbf{Execution Time (min)} \\
        \midrule
        \textbf{chr12a} 
            & VQE         & 144 & 163 & 2443& 715 & 1728 & 1917.5 \\
            & CVaR VQE    & 144 & 163 & 2443& 715 & 1728 & 777.35 \\
            & PCE         & 11 & 528 & 4048 & 880 & 924 & 537.64 (470.78 + 66.86) \\
        \midrule
        \textbf{chr12b} 
            & VQE         & 144 & 163 & 2443& 715 & 1728 & 1927.9 \\
            & CVaR VQE    & 144 & 163 & 2443& 715 & 1728 & 772.45 \\
            & PCE         & 11 & 528 & 4048 & 880 & 924 & 592.71 (521.41 + 71.3) \\
        \midrule
        \textbf{chr12c} 
            & VQE         & 144 & 163 & 2443& 715 & 1728 & 1910.4 \\
            & CVaR VQE    & 144 & 163 & 2443& 715 & 1728 &760.23 \\
            & PCE         & 11 & 528 & 4048 & 880 & 924 & 930.45 (880.15 + 50.3) \\
        \midrule
        \textbf{nug12} 
            & VQE         & 144 & 163 & 2443& 715 & 1728 &  1947.8 \\
            & CVaR VQE    & 144 & 163 & 2443& 715 & 1728 &  765.55 \\
            & PCE         & 11 & 528 & 4048 & 880 & 924 & 1169.12 (1035.5 + 133.61) \\
        \midrule
        \textbf{rou12} 
            & VQE         & 144 & 163 & 2443& 715 & 1728 & 1961.6 \\
            & CVaR VQE    & 144 & 163 & 2443& 715 & 1728 & 796.50 \\
            & PCE         & 11 & 528 & 4048 & 880 & 924 & 1500.43 (1323.9 + 176.5)  \\
        \midrule
        \textbf{scr12} 
            & VQE        & 144 & 163 & 2443& 715 & 1728 & 1906.6 \\
            & CVaR VQE    & 144 & 163 & 2443& 715 & 1728 & 783.40 \\
            & PCE         & 11 & 528 & 4048 & 880 & 924 & 831.38 (732.18 + 99.2) \\
        \midrule
        \textbf{tai12a} 
            & VQE         & 144 & 163 & 2443& 715 & 1728 & 1883.8 \\
            & CVaR VQE    & 144 & 163 & 2443& 715 & 1728 & 755.45 \\
            & PCE         & 11 & 528 & 4048 & 880 & 924 & 1545.33 (1362.9 + 182.43) \\
        \midrule
        \textbf{tai12b} 
            & VQE        & 144 & 163 & 2443& 715 & 1728 & 1911.3 \\
            & CVaR VQE    & 144 & 163 & 2443& 715 & 1728 & 788.96 \\
            & PCE         & 11 & 528 & 4048 & 880 & 924 & 1206.2 (1072.1 + 134.1) \\
        \bottomrule
    \end{tabular*}
    \label{tab:resource_usage_qap}
\end{table*}

\begin{table*}[p]
    \caption{Resource Usage by Instance and Approach. Each instance's resources are detailed for VQE, CVaR VQE, and PCE approaches. In case of PCE Total Time (Optimization time + Post Processing Time), since it uses bit swap search as a post processing strategy.}
    \renewcommand{\arraystretch}{1.3}
    \centering
    \footnotesize
    \begin{tabular*}{\textwidth}{@{\extracolsep{\fill}} lccccccc}
        \toprule
        \textbf{Instance} & \textbf{Approach} & \textbf{Qubits} & \textbf{Depth} & \textbf{Gate Count} & \textbf{2-Qubit Gates} & \textbf{Parameters} & \textbf{Execution Time (min)} \\
        \midrule
        \textbf{2x10 S0} 
            & VQE         & 50 & 69 & 845 & 245 & 600 & 2135.68 \\
            & CVaR VQE    & 50 & 69 & 845 & 245 & 600 & 1727.78 \\
            & PCE         & 7 & 336 & 1568 & 336 & 364 & 55.39 (54.21 + 1.18) \\
        \midrule
        \textbf{2x10 S1} 
            & VQE         &  46 & 65 & 777 & 225 & 552 & 1680.04 \\
            & CVaR VQE    & 46 & 65 & 777 & 225 & 552 & 1343.81 \\
            & PCE         & 7 & 336 & 1568 & 336 & 364 & 47.35 (46.48 + 0.87) \\
        \midrule
        \textbf{3x20 S0} 
            & VQE         & 84 & 103 & 1423 & 415 & 1008 & 2050.56 \\
            & CVaR VQE    & 84 & 103 & 1423 & 415 & 1008 & 1358.66 \\
            & PCE         & 8 & 384 & 2080 & 448 & 480 & 164.90 (155.35 + 9.56) \\
        \midrule
        \textbf{3x20 S1} 
            & VQE         & 80 & 99 & 1355 & 395 & 960 & 2494.05 \\
            & CVaR VQE    &  80 & 99 & 1355 & 395 & 960 & 1857.64 \\
            & PCE         & 8 & 384 & 2080 & 448 & 480 & 155.55 (147.6 + 7.95) \\
        \midrule
        \textbf{4x30 S0} 
            & VQE         & 118 & 137 & 2001 & 585 & 1416 & 4283.23 \\
            & CVaR VQE    & 118 & 137 & 2001 & 585 & 1416 & 3629.8 \\
            & PCE         & 10 & 480 & 3320 & 720 & 760 & 504.55 (460.6 + 43.95) \\
        \midrule
        \textbf{4x30 S1} 
            & VQE         & 118 & 137 & 2001 & 585 & 1416 & 4663.91 \\
            & CVaR VQE    & 118 & 137 & 2001 & 585 & 1416 & 3712.6 \\
            & PCE         & 10 & 480 & 3320 & 720 & 760 & 512.22 (469.28 + 44.93)\\
        \midrule
        \textbf{5x40 S0} 
            & VQE         & 154 & 165 & 1691 & 459 & 1232 & 6956.9 \\
            & CVaR VQE    & 154 & 165 & 1691 & 459 & 1232 & 5958.5 \\
            & PCE         & 11 & 528 & 4048 & 880 & 924 & 1029.16 (887.2 + 141.96) \\
        \midrule
        \textbf{5x40 S1} 
            & VQE         & 154 & 165 & 1691 & 459 & 1232 & 6693.63 \\
            & CVaR VQE    & 154 & 165 & 1691 & 459 & 1232 & 5719.3 \\
            & PCE         & 11 & 528 & 4048 & 880 & 924 & 1020.86 (896.80 + 124.06)\\
        \bottomrule
    \end{tabular*}
    \label{tab:resource_usage_ms}
\end{table*}

\end{document}